\documentclass[aps,pre,showkeys,showpacs,
reprint,
longbibliography,
floatfix]{revtex4-1}

\usepackage[utf8]{inputenc}
\usepackage[T1]{fontenc}
\usepackage{amsmath}
\usepackage{graphicx}
\usepackage{subcaption}
\usepackage{psfrag}
\usepackage{xcolor}
\usepackage{listings}
\usepackage[colorlinks=true,hyperfootnotes=true,breaklinks=true]{hyperref}

\usepackage[justification=centerlast]{caption}  
\newcommand{\varname}[1]{\texttt{#1}}   
\newcommand{\procname}[1]{\textsc{#1}}  

\begin{document}
\title{Efficient space virtualisation for Hoshen--Kopelman algorithm}
\author{M.~Kotwica}
\author{P.~Gronek}
\author{K.~Malarz}
\homepage{http://home.agh.edu.pl/malarz/}
\email{malarz@agh.edu.pl}
\affiliation{\href{http://www.agh.edu.pl/}{AGH University of Science and Technology},
\href{http://www.pacs.agh.edu.pl/}{Faculty of Physics and Applied Computer Science},
al. Mickiewicza 30, 30-059 Krak\'ow, Poland}

\begin{abstract}
In this paper the efficient space virtualisation for the Hoshen--Kopelman algorithm is presented.
We observe minimal parallel overhead during computations, due to negligible communication costs.
The proposed algorithm is applied for computation of random-site percolation thresholds for four dimensional simple cubic lattice with sites' neighbourhoods containing next-next-nearest neighbours (3NN).
The obtained percolation thresholds are
$p_C(\text{NN})=0.19680(23)$,
$p_C(\text{2NN})=0.08410(23)$,
$p_C(\text{3NN})=0.04540(23)$,
$p_C(\text{2NN+NN})=0.06180(23)$,
$p_C(\text{3NN+NN})=0.04000(23)$,
$p_C(\text{3NN+2NN})=0.03310(23)$,
$p_C(\text{3NN+2NN+NN})=0.03190(23)$,
where 2NN and NN stand for next-nearest neighbours and nearest neighbours, respectively.
\end{abstract}



\date{\today}
\maketitle

\section{Introduction}

Percolating systems~\cite{Broadbent1957,*Frisch1961,*Frisch1962,bookDS,bookBB,bookHK,bookMS} are examples of system where purely geometrical phase transition may be observed (see Ref.~\cite{SABERI20151} for recent review).
The majority of percolating systems which may be mapped to real-world problems deal with two- or three-dimensional space and ranges from condensed matter physics \cite{PhysRevB.89.054409,*Silva2011,*Shearing2010,*Halperin2010} via rheology \cite{Mun2014,*Amiaz2011,*Bolandtaba2011,*Mourzenko2011} and forest fires \cite{Abades2014,*Camelo-Neto2011,*Guisoni2011,*Simeoni2011,*Kaczanowska2002} to immunology \cite{Silverberg2014,*Suzuki2011,*Lindquist2011,*Naumova2008,*Floyd2008} and quantum mechanics \cite{Chandrashekar2014}.
However, computer simulations are conducted also for systems with non-physical dimensions $d$ (up to $d=13$)~\cite{Stauffer2000,PhysRevE.64.026115,PhysRevE.67.036101,Ballesteros1997,Marck1998,0305-4470-31-15-010}.

In classical approach only the nearest neighbours (NN) of sites in $d$-dimensional system are considered.
However, complex neighbourhoods may have both, theoretical \cite{ISI:000416616200002,ISI:000414461700008} and practical \cite{ISI:000246890600043,ISI:000314677100022,Masin,ISI:000380097300006,ISI:000404810700001,ISI:000361674600003}
applications.
These complex neighbourhoods may include not only NN but also next-nearest neighbours (2NN) and next-next-nearest neighbours (3NN).

One of the most crucial feature describing percolating systems is a {\em percolation threshold} $p_C$.
In principle, this value separates two phases in the system; 
\begin{itemize}
\item if sites are occupied with probability $p<p_C$ the system behaves as `an insulator',
\item while for $p>p_C$ the system exhibit attributes of `a~conductor'.
\end{itemize}
Namely, for $p=p_C$ the giant component containing most of occupied sites appear for the first time.
The cluster of occupied sites spans from the one edge of the system to the other one (both being ($d-1$)-dimensional hyper-planes).
This allows for direct flow of material (or current) from one to the other edge of the systems.
For $p>p_C$ this flow is even easier while for $p<p_C$ the gaps of unoccupied (empty) sites successfully prevent such flow \footnote{The terminology of material flow between edges of the system comes from the original papers introducing the percolation term \cite{Broadbent1957,*Frisch1961,*Frisch1962} in the subject of rheology.}.

Among so far investigated systems also four dimensional systems were considered~\cite{Ballesteros1997,Marck1998,0305-4470-31-15-010,PhysRevE.64.026115,PhysRevE.67.036101}.
The examples of percolation thresholds for four dimensional lattices and NN neighbours are presented in Tab.~\ref{tab-PT-ref}.

\begin{table}[!htbp]
\caption{\label{tab-PT-ref} The critical values of $p_C$ for various four dimensional lattices with NN neighbours and various site coordination number $z$.}
\begin{ruledtabular}
\begin{tabular}{lrlr}
lattice  &$z$& $p_C$          & Ref. \\ \hline
diamond  & 5  & 0.2978(2)     & \cite{Marck1998}\\
SC       & 8  & 0.196901(5)   & \cite{Ballesteros1997}\\
SC       & 8  & 0.196889(3)   & \cite{PhysRevE.64.026115}\\
SC       & 8  & 0.1968861(14) & \cite{PhysRevE.67.036101}\\
Kagom\'e & 8  & 0.2715(3)     & \cite{0305-4470-31-15-010}\\	
BCC 	 & 16 & 0.1037(3)     & \cite{Marck1998}\\
FCC 	 & 24 & 0.0842(3)     & \cite{Marck1998}\\
\end{tabular}
\end{ruledtabular}
\end{table}

In this paper we
\begin{itemize}
\item propose an efficient space virtualisation for Hoshen--Kopelman algorithm \cite{Hoshen1976a} employed for occupied sites clusters labelling,
\item estimate the percolation thresholds for a four dimensional simple cubic (SC) lattice with complex neighbourhoods, i.e. neighbourhoods containing various combinations of NN, 2NN and 3NN neighbours.
\end{itemize}

{While the Hoshen--Kopelman method is good for many problems---}
{for instance, the} parallel version of {the} Hoshen--Kopelman algorithm has been successfully applied for lattice-Boltzmann simulations \cite{FRIJTERS201592}{---probably it is not the best method available to find the thresholds. One can even grow single clusters by a Leath type of algorithm} \cite{Leath1976}{, and find the threshold where the size distribution is power-law, as has been done in many works in three dimensions. For high-dimensional percolation, both Grassberger} \cite{PhysRevE.67.036101}{, and Mertens and Moore} \cite{PhysRevE.98.022120} {use a method where you do not even have a lattice, but make a list of the coordinates of all the sites that have been visited, and using computer-science type of structures (linked lists and trees, etc.) one can search if a site has already been visited in a short amount of time.  Mertens and Moore} \cite{PhysRevE.96.042116} {have also recently proposed an intriguing method where they use basically invasion percolation (along with the various lists) to grow large clusters that self-organize to the critical point.  Both groups have gone up to 13 dimensions using these methods.}

\section{\label{sec-approach}Methodology}

To evaluate the percolation thresholds the finite-size scaling technique~\cite{Fisher1971,bookVP,Binder1992,bookDL} has been applied.
According to this theory the quantity $X(p)$ characterising the system in the vicinity of critical point $p_C$ scales with the system linear size $L$ as 
\begin{equation}
\label{eq-scaling}
X(p;L) = L^{-x}\cdot\mathcal{F}\left((p-p_C)L^{1/\nu}\right),
\end{equation}
where $\mathcal{F}(\cdot)$ is a scaling function, $x$ is a scaling exponent and $\nu$ is a critical exponent associated with the correlation length \cite{bookDS}.

For $p=p_C$ the value of $L^xX(p;L)=\mathcal{F}(0)$ does not depend on the system linear size $L$ which allows for predicting the position of percolation threshold $p_C$ as curves $L^xX(p;L)$ plotted for various values of $L$ cross each other at $p=p_C$.
Moreover, for appropriate selection also the value of critical exponent $\nu$ the dependencies $L^xX(p;L)$ vs. $\left((p-p_C)L^{1/\nu}\right)$ collapse into a single curve independently on $L$.

\begin{figure*}
\psfrag{L=}[][c]{{\small $L=$}}
\psfrag{p}{$p$}
\psfrag{W}[][c]{$W(p;L)$}
	\begin{subfigure}[b]{0.330\textwidth}
		\caption{NN\label{fig-NN}}
		\includegraphics[width=\textwidth]{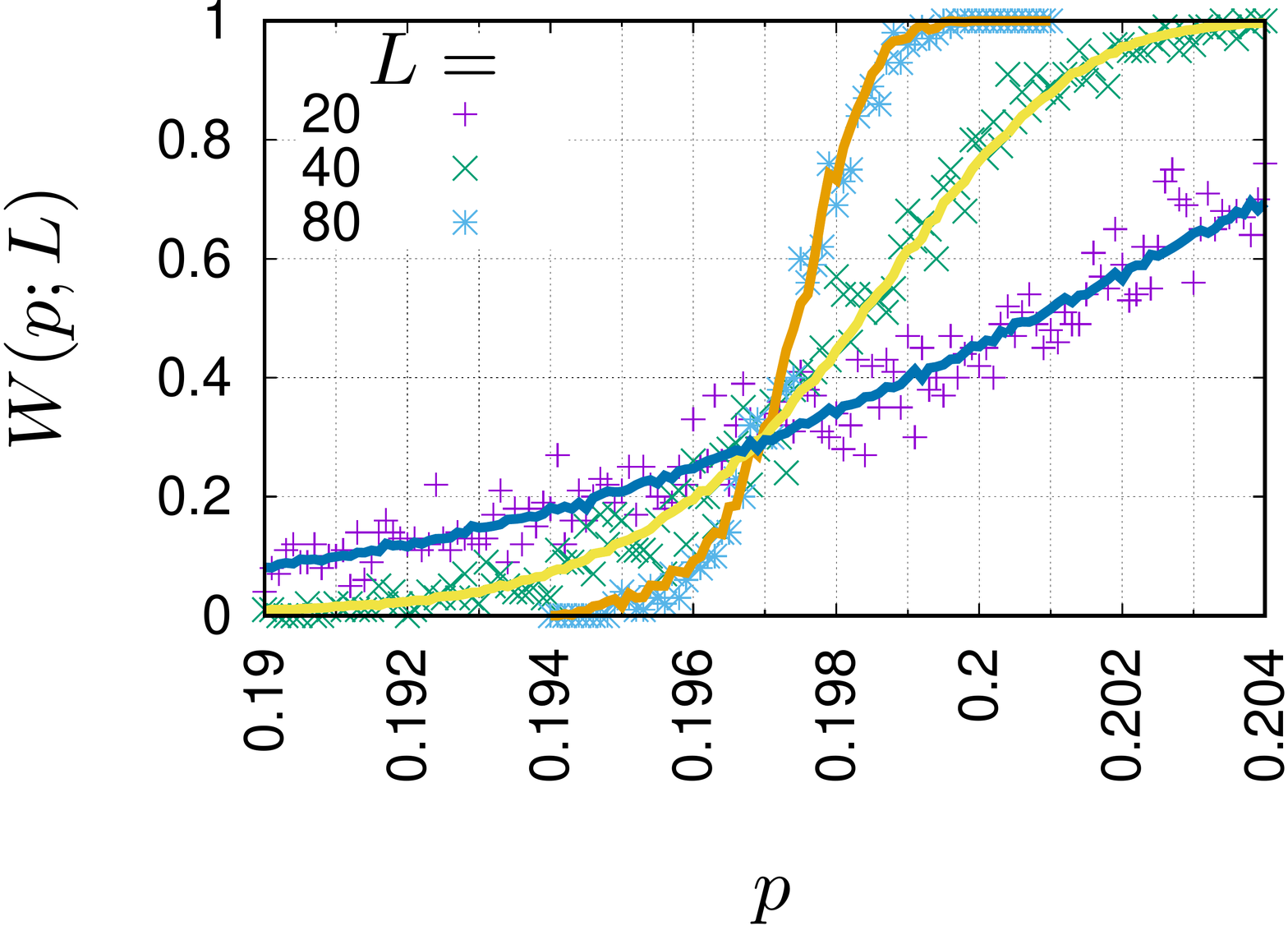}
	\end{subfigure}\\
	\begin{subfigure}[b]{0.329\textwidth}
	\caption{2NN\label{fig-2NN}}
	{\includegraphics[width=\textwidth]{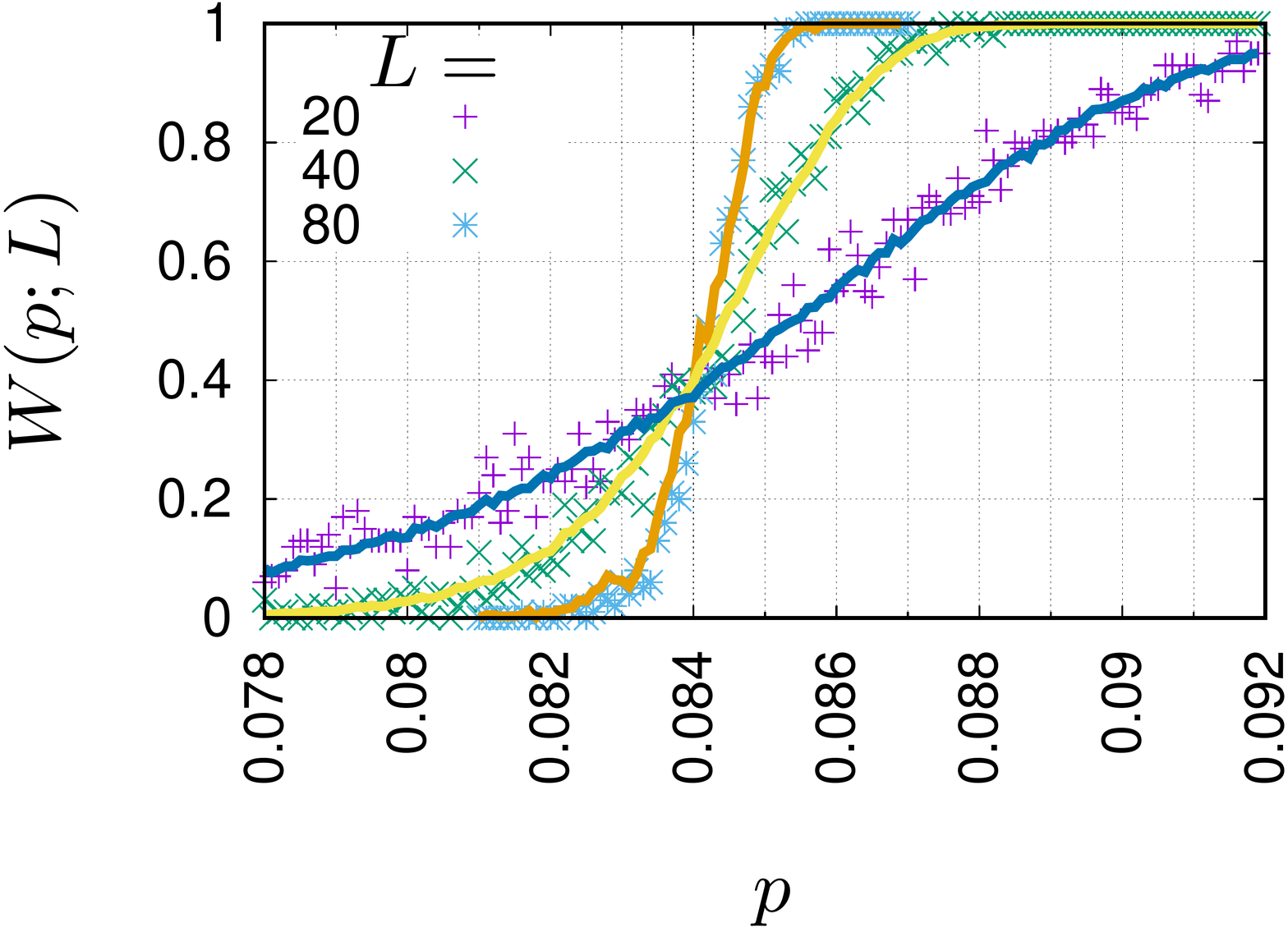}}
	\end{subfigure}
\hfill
	\begin{subfigure}[b]{0.329\textwidth}
	\caption{2NN+NN\label{fig-2NN_NN}}
	{\includegraphics[width=\textwidth]{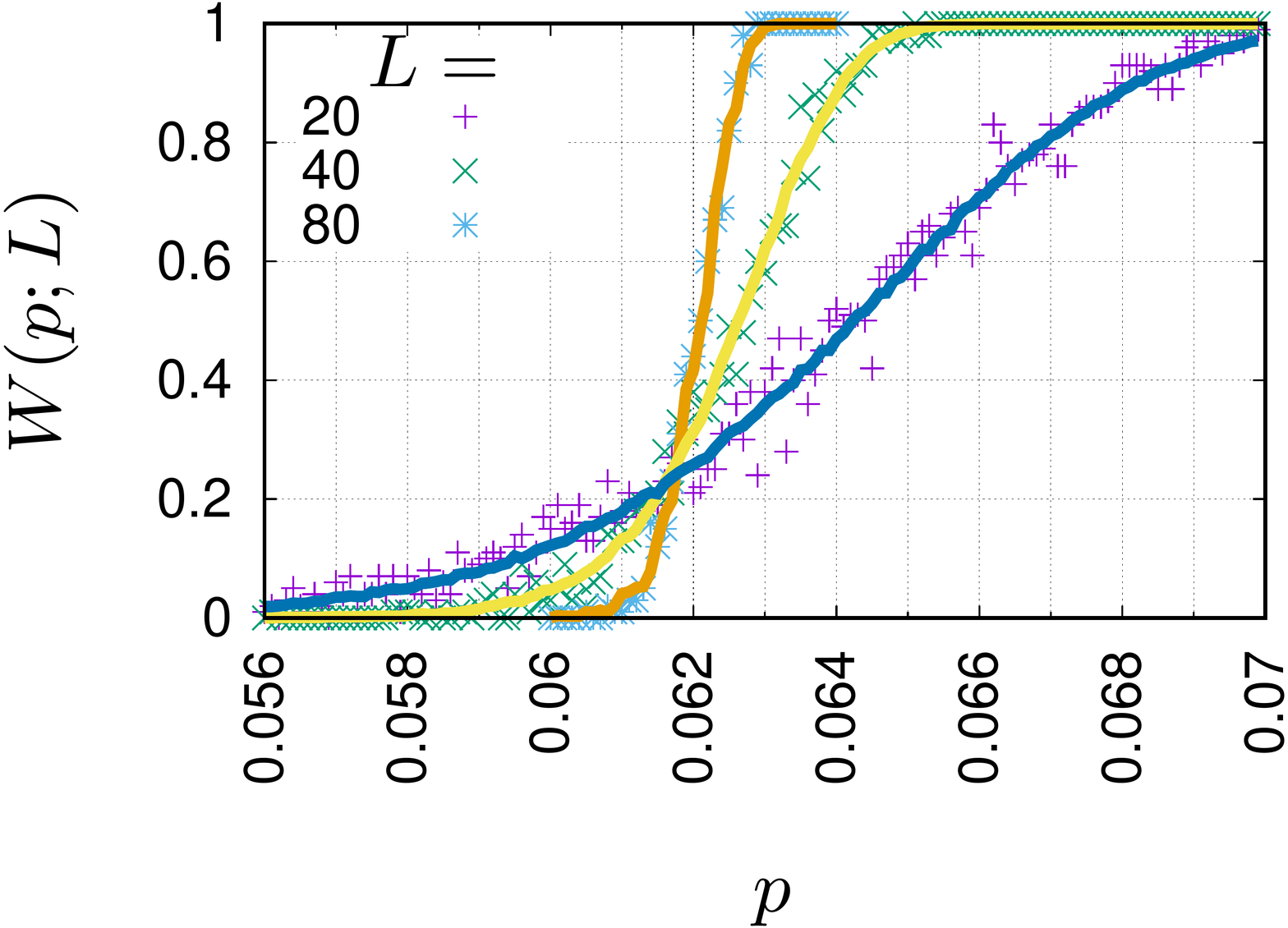}}
	\end{subfigure}
\hfill
	\begin{subfigure}[b]{0.329\textwidth}
	\caption{3NN\label{fig-3NN}}
	{\includegraphics[width=\textwidth]{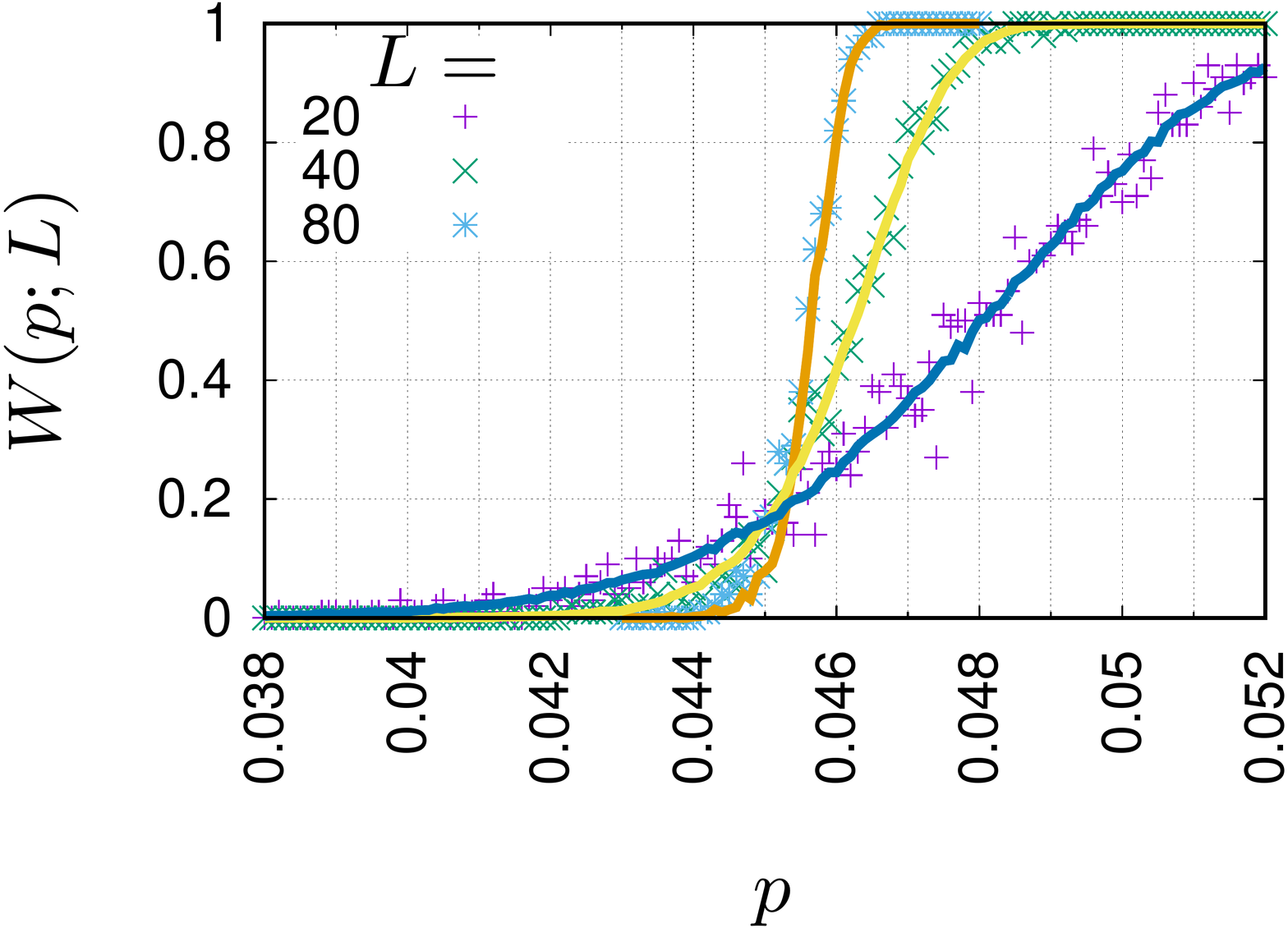}}
	\end{subfigure}\\
	\begin{subfigure}[b]{0.329\textwidth}\caption{3NN+NN\label{fig-3NN_NN}}{\includegraphics[width=\textwidth]{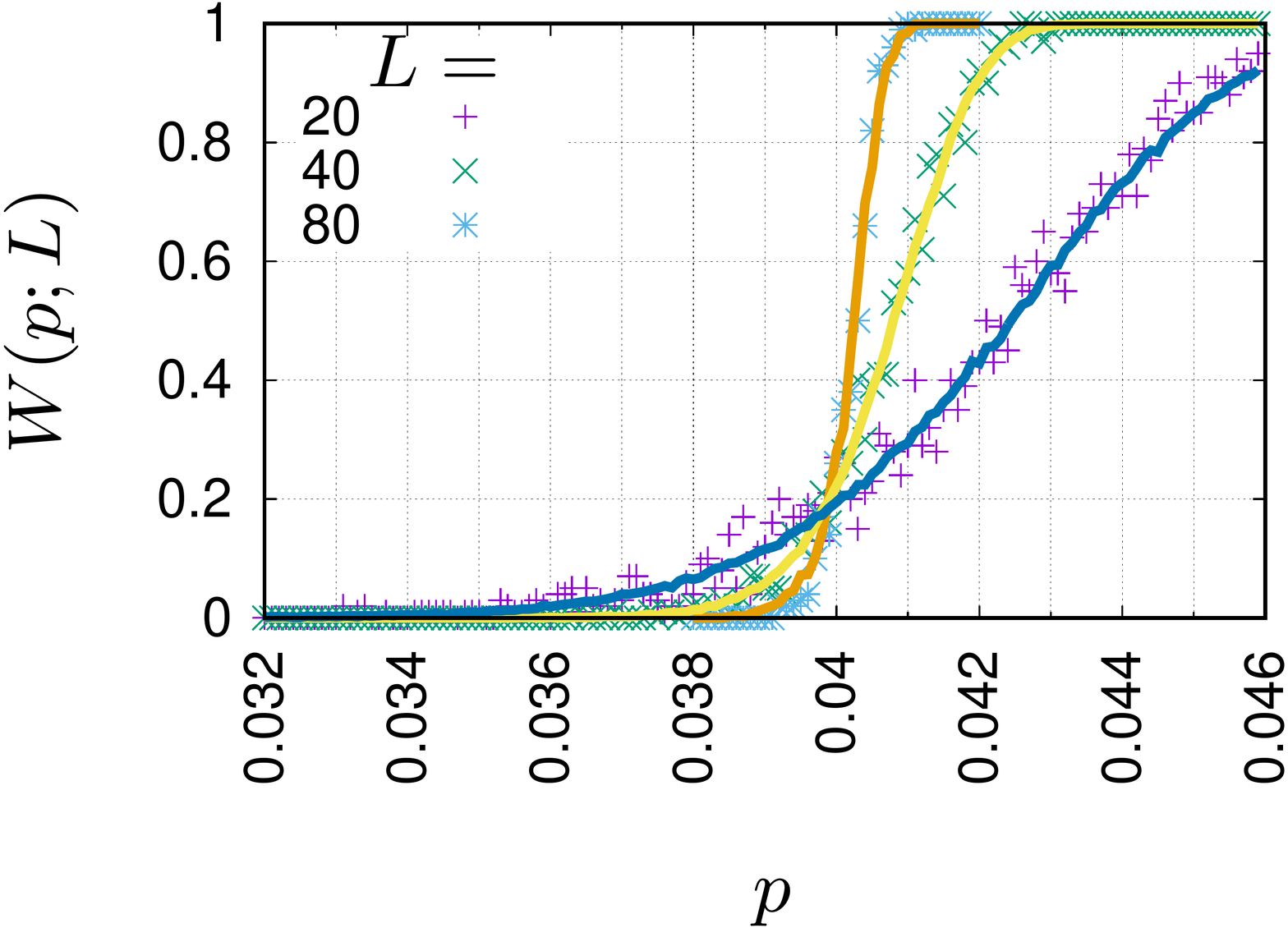}}\end{subfigure}
\hfill
	\begin{subfigure}[b]{0.329\textwidth}\caption{3NN+2NN\label{fig-3NN_2NN}}{\includegraphics[width=\textwidth]{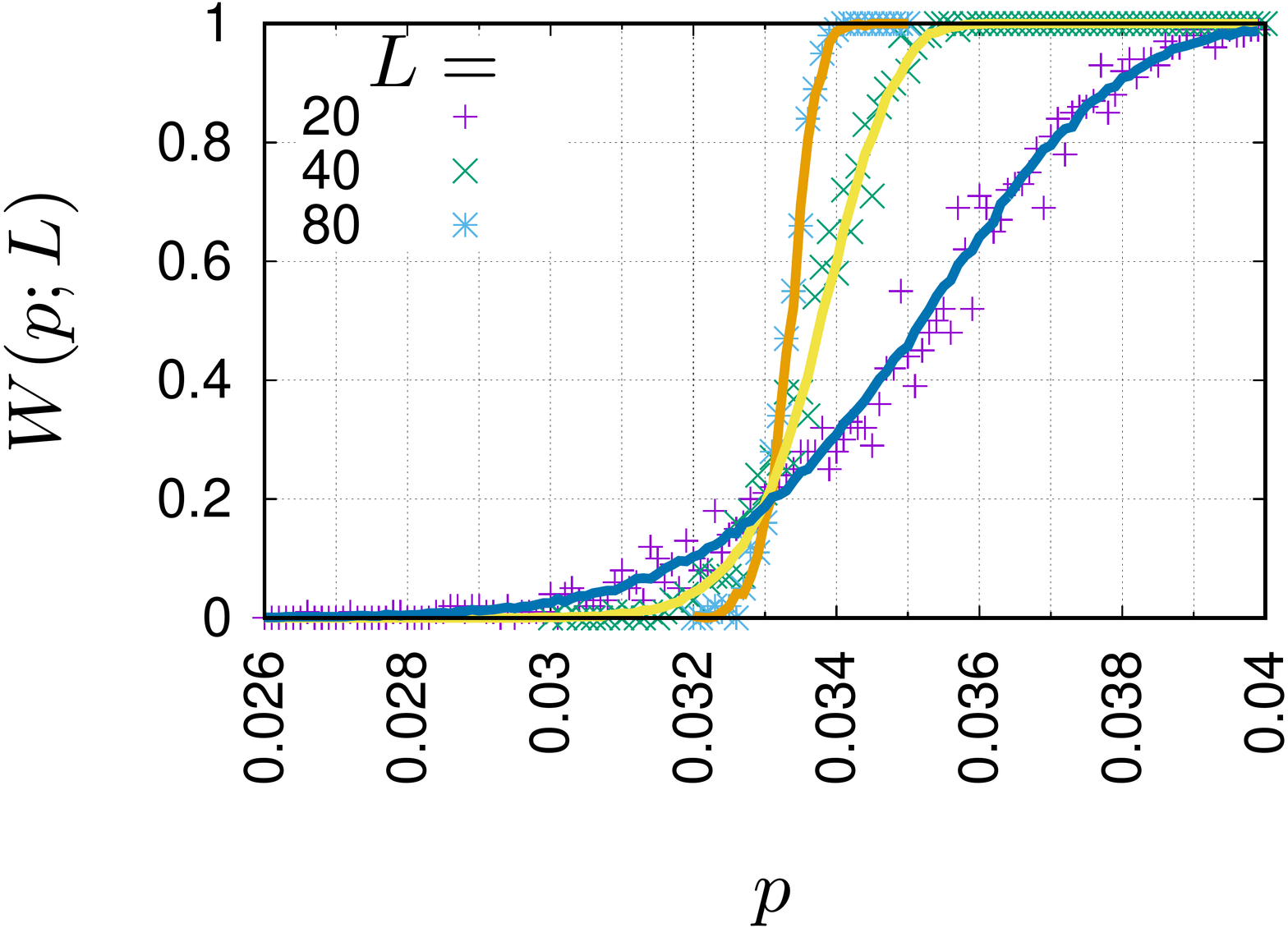}}\end{subfigure}
\hfill
	\begin{subfigure}[b]{0.329\textwidth}\caption{3NN+2NN+NN\label{fig-3NN_2NN_NN}}{\includegraphics[width=\textwidth]{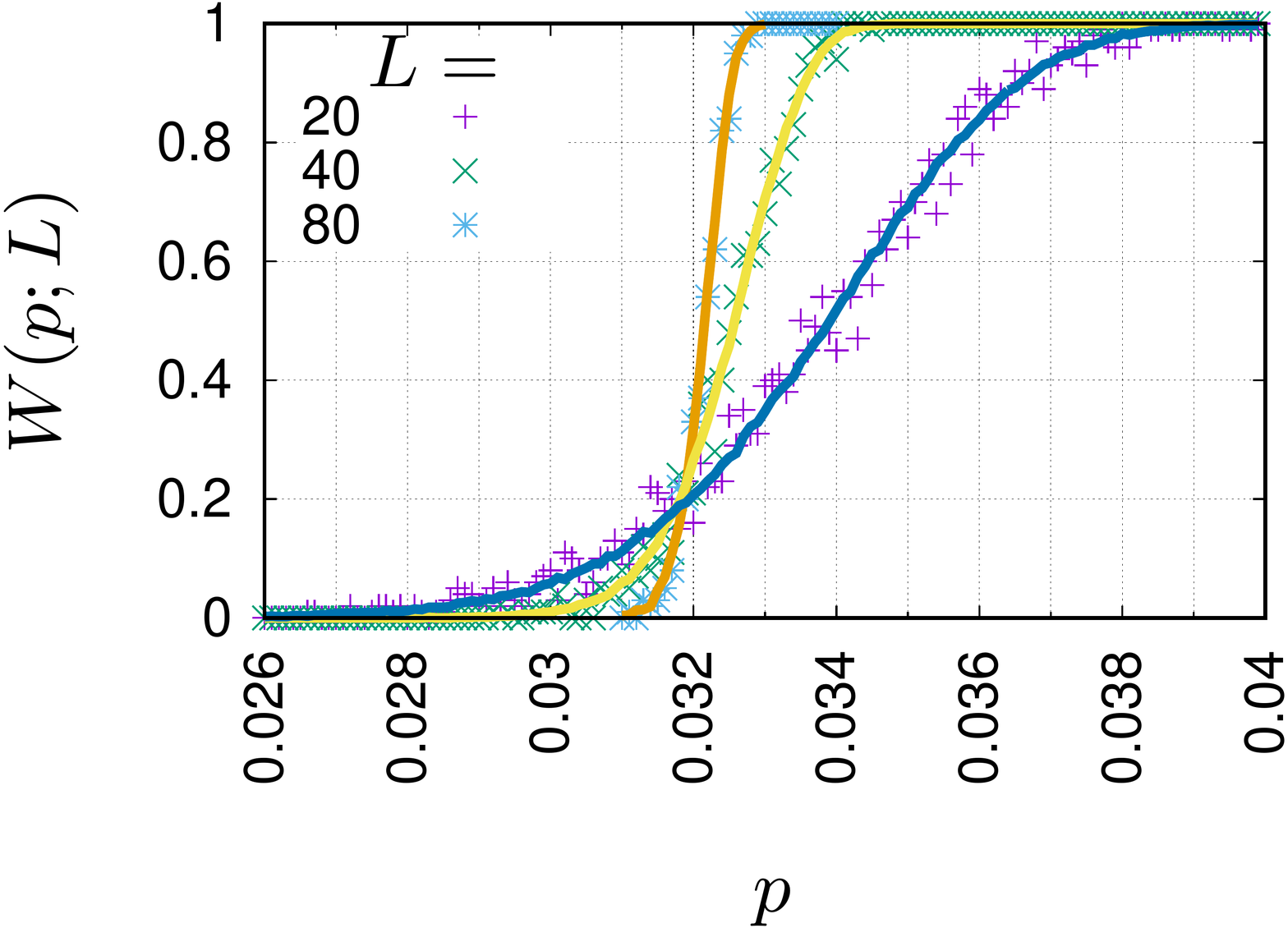}}\end{subfigure}
\caption{\label{fig-W} (Colour online).
Wrapping probability $W(p;L)$ vs. occupation probability $p$.
The results are averaged over $R=10^2$ (symbols) or $R=10^4$ (lines) simulations.}
\end{figure*}

Such technique encounters however one serious problem.
Namely, numerically deduced curves $L^xX(p;L)$ for various $L$ {\em rather seldom} cross each other in a single point, particularly when the number of independent simulations is not huge (see Fig.~\ref{fig-W}, where examples of dependencies $L^xX(p;L)$ for $L=20$, 40, 80 and $R=10^2$ (symbols) and $R=10^4$ (lines) are presented).

The remedy for this troubles has been proposed by Bastas et al.~\cite{PhysRevE.84.066112,Bastas2014} and even simplified in Ref.~\cite{Malarz2015}.
The methodology proposed in Ref.~\cite{Malarz2015} allows for estimation of percolation threshold $p_C$ also for relatively low sampling if the quantity $X(p;L)$ is chosen smartly. 
Namely, as $X(p;L)$ should be chosen quantity for which scaling exponent $x$ is equal to zero. 
One of such quantity is the wrapping probability 
\begin{equation}
\label{eq-W}
W(p;L)=N(p;L)/R
\end{equation} 
describing fraction of percolating lattices among $R$ lattices constructed for $pL^d$ occupied sites and fixed values of $p$ and $L$, where $d$ is a geometrical space dimension and $N(p;L)$ is a number of percolating lattices.

According to Refs.~\cite{PhysRevE.84.066112,Bastas2014,Malarz2015} instead of searching common crossing point of $L^xX(p;L)$ curves  for various $L$ one may wish to minimise
\begin{equation}
\label{eq-lambda}
\lambda(p)\equiv\sum_{i\ne j}\left[H(p;L_i)-H(p;L_j)\right]^2,
\end{equation}
where
\begin{equation}
H(p;L)\equiv W(p;L) + 1/W(p;L).
\end{equation}
The minimum of $\lambda(p)$ near `crossing points' of $W(p;L)$ curves plotted for various sizes $L$ yields the estimation of percolation threshold $p_C$.

Such strategy allowed for estimation of percolation thresholds $p_C$ for simple cubic lattice ($d=3$) with complex neighbourhoods (i.e. containing up to next-next-next-nearest neighbours) with relatively low-sampling ($R=10^4$) \cite{Malarz2015}.
Unfortunately, reaching similar accuracy as in Ref.~\cite{Malarz2015} for similar linear sizes of the system and for increased space dimension ($d=4$) requires increasing sampling by one order of magnitude (to $R=10^5$).
This however makes the computations times extremely long.
In order to overcome this trouble we propose efficient way of problem parallelisation.

\section{\label{sec-compu}Computations}

Several numerical techniques allow for clusters of connected sites identification~\cite{Hoshen1976a,Leath1976,Newman2001,Torin2014}.
Here we apply the Hoshen--Kopelman algorithm \cite{Hoshen1976a}, which allows for sites labelling in a such way, that occupied sites in the same cluster have assigned the same labels and different clusters have different labels associated with them.

The simulations were carried out on Prometheus~\cite{pro}, an Academic Computer Centre Cyfronet AGH-UST operated parallel supercomputer, based on Hewlett--Packard Apollo 8000 Gen9 technology.
It consists of 5200 computing 2232 nodes, each with dual 12-core Xeon E5-2680v3 CPUs,  interconnected by Infiniband FDR network and over 270 TB of storage space.
It supports a wide range of parallel computing tools and applications, including MVAPICH2~\cite{MVAPICH} MPI implementation for C and Fortran compilers and provides 2.4 PFlops of computing performance, giving it 77th position on November 2017 edition of Supercomputer Top500 list~\cite{TOP500}.

\subsection{\label{sec-implementation} Implementation}

\begin{figure}
	\centering
	\begin{subfigure}[b]{0.5\textwidth}
		\caption{in-domain\label{fig-sep-dep}}
		\includegraphics[height=3cm]{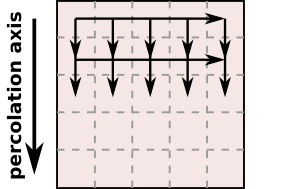}
	\end{subfigure}\\
	\begin{subfigure}[b]{0.5\textwidth}
		\caption{inter-domain\label{fig-sub-domains}}
		\includegraphics[height=3cm]{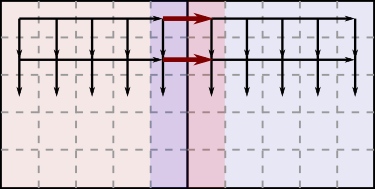}
	\end{subfigure}
  \caption{\label{fig-domains} (Colour online).  Parallel subdomains 2-$d$ dependencies model:
      in-domain across slice (thin horizontal arrows),
      in-domain between slices (thin vertical arrows),
      inter-domain (bold horizontal arrows).}
\end{figure}

One of the problems encountered is high memory size complexity of $\mathcal{O}(L^4)$, resulting from space being a hyper-cube growing in each direction. The memory limit on the machines the program was run did not allow $L \gg 120$. Several solutions were put in consideration, one of which was splitting single simulations' calculations between nodes. In the classical version of the algorithm there are sequential dependencies both subsequent 3-$d$ slices perpendicular to the axis of percolation and across any given such slice (see Fig.~\ref{fig-sep-dep}).

If speed is to be ignored, sequential dependencies could be simply distributed across the domains. Then a domain succeeding another one in any direction (e.g. blue one succeeding red one on Fig.~\ref{fig-sub-domains}) should be sent information on the class of sites in the area of touch ($L^3$ in size) as well as synchronise aliasing arrays. It could be done slice-wise or domain-wise. Slice-wise approach minimises aliasing desynchronisation while domain-wise concentrates communication.

Work between domains can be parallelised. If any domain finds a percolating cluster, it also percolates in the whole hyper-cube, if not, synchronisation of aliasing and both first and last slice cluster ID lists. It requires either totally rebuilding label array or costly label identity checks.

Because of the aforementioned costs and complications, another approach was chosen. Parallelization was only used to accelerate calculations (see Sec.~\ref{sec-efficiency}) while the lack of memory problem was solved using space virtualisation.

\begin{figure}
	\begin{subfigure}[b]{0.5\textwidth}
	\caption{Clustering over adjacent slices\label{subfig-adj-sli}}
	{{\includegraphics[width=\textwidth]{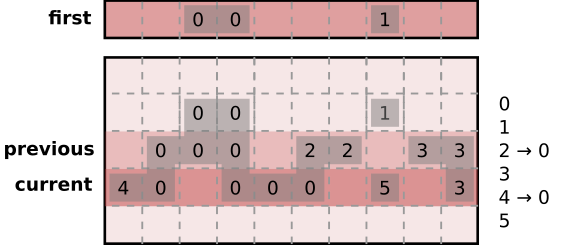}}}\end{subfigure}\\
	\begin{subfigure}[b]{0.5\textwidth}
	\caption{Virtualisation step\label{subfig-sli-at-end}}
	{{\includegraphics[width=\textwidth]{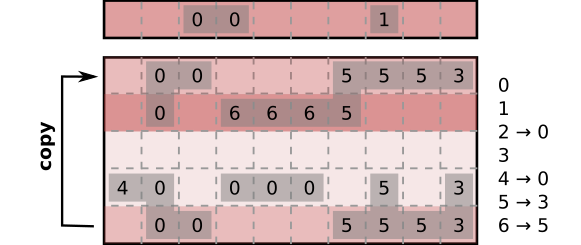}}}\end{subfigure}
  \caption{\label{fig-buffering} (Colour online).
  Hyperspace buffering in virtualisation (2-$d$ model).
  Total cost: $\mathcal{O}(L^4)$, like without buffering.}
\end{figure}

\subsubsection{\label{sec-mpi} Message Passing Interface}
Message Passing Interface (MPI)~\cite{MPI2014,*AdvancedMPI2014} is prevailing model of parallel computation on distributed memory systems, including dedicated massively parallel processing supercomputers and cluster systems.
It is constructed as a library interface, to be integrated with computer programs written in Fortran or C languages.
The main advantage of MPI is its portability and ease of use. Software vendors are presented with clearly defined set of routines that can be implement efficiently, with hardware support provided for specific computer architectures.
Several well-tested open-source implementations are also available~\cite{MPICH}. The development of MPI specification is maintained by MPI Forum~\cite{MPIforum}, an academia and industry-based consortium, led by University of Tennessee, Knoxville, TN, USA.

\subsubsection{\label{sec-space-virtualization} Space virtualisation}

Space virtualisation is possible due to the fact there are very limited information that need to be extracted from the hyper-cube, namely: does it contain at least one percolating cluster. However, any percolating cluster is also percolating for a 4-$d$ slice of any depth perpendicular to the percolation axis, including any cut of depth two. Indeed, clusterising any 3-$d$ slice across the percolation axis requires its immediately preceding slice to be fully available as well as clusterised (see Fig.~\ref{fig-sep-dep}). It implies only three slices are essentially needed for calculations: the current one, the previous one and the first one, represented on Fig.~\ref{subfig-adj-sli}).

Any iteration over slice buffer introduces additional cost of copying the last slice as the next slice's immediate predecessor. Due to that, minimal buffer of depth three is highly sub-optimal. For any buffer depth $D$, the additional cost of a single iteration over buffer is $\mathcal{O}(L^3)$ and $\left\lfloor {L/D} \right\rfloor$ such iterations are needed, for the total cost of buffering being $\mathcal{O}({L^4}/{D})$, which implies $D$ should be as big as possible.

In fact, buffering can even benefit performance in some cases as smaller chunk of memory may be possible to fit in cache along with label aliasing array. If so, the cost of copying can be more than compensated by massive reduction of number of memory accesses.

\subsubsection{\label{sec-parallelism} Parallelism}

Because calculations are performed sequentially along the percolation axis, sequentially both over a slice and between slices, no asymptotic speed up is gain from that. However, for each state occupation probability $p$, many simulations ($R$) are run for the results to be meaningful, which leads to the total cost of $\mathcal{O}(L^4 R)$.

As the simulations are fully independent and only their results are to be combined, the program can be speed-up by utilising parallelism over tasks. The only communication needed is collecting the results (whenever a percolating cluster was found or not) at the end of calculations, which is close to $\mathcal{O}(1)$. Theoretical cost is then $\mathcal{O} \left( {L^4 R}/{N} \right)$, where $N$ is the number of computational nodes.

In practice it is unnecessary to map each task to a separate native process, which would require huge amounts of CPU cores (thousands to hundreds of thousands). However, each simulation has a similar execution time, which means no run-time work re-balancing is needed so MPI processes can be used with tasks distributed equally among them. Optimally, the number of processes should be a divisor of $N$.

Utilising more than one process per node puts additional limit on memory, which implies reduction of virtualisation buffer depth. For $N$ nodes and $C$-core architecture that is:
\begin{equation}
D_2 = D_1/C.
\end{equation}

This operation implies additional cost of roughly multiplying buffering costs $C$ times, while reducing all costs $C$ times due to multiplying the total number of tasks, which is a clear advantage. Because of that, every core is assigned a separate process.

Threading could be used within a single node but due to close to no communication between tasks, it would make the code more sophisticated with very little performance advantage (only reducing $\mathcal{O}(1)$ cost of communication).

\section{\label{sec-results}Results}

\subsection{\label{sec-efficiency}Speed-up and efficiency}

One of  the most frequently used performance metric of parallel processing is speedup \cite{Wu1999,Scott2005}. 
Let $\tau(L^d,1)$ be the execution time of the sequential algorithm and let $\tau(L^d,\mathcal{N})$ be the execution of the parallel algorithm executed on $\mathcal{N}$ processors, where $L^d$ is the size of the computational problem.
A speed-up of parallel computation is defined as
\begin{equation}
	\mathcal{S}(\mathcal{N})=\frac{\tau(L^d,1)}{\tau(L^d,\mathcal{N})}
\end{equation}
the ratio of the sequential execution time to the parallel execution time.
One would like to achieve $S(\mathcal{N})=\mathcal{N}$, so called perfect speed-up~\cite{Amdahl1967}.
In this case the problem size stays fixed but the number of processing elements are increased.
This are referred to as strong scaling~\cite{Scott2005}.
In general, it is harder to achieve good strong-scaling at larger process counts since the communication overhead for many/most algorithms increases in proportion to the number of processes used.

Another metric to measure the performance of a parallel algorithm is efficiency, $\mathcal{E}$, defined as
\begin{equation}
	\mathcal{E}(\mathcal{N})=\frac{\mathcal{S}(\mathcal{N})}{\mathcal{N}}.
\end{equation}
Speedup and efficiency are therefore equivalent measures, differing only by the constant factor $\mathcal{N}$. 

\begin{figure}
\psfrag{x}[r][]{$\mathcal{S}(\mathcal{N})=\mathcal{N}$}
\psfrag{a}[r][]{$p=p_C$: NN}
\psfrag{b}[r][]{3NN+2NN+NN}
\psfrag{S}{$\mathcal{S}$}
\psfrag{N}{$\mathcal{N}$}
\includegraphics[width=0.45\textwidth]{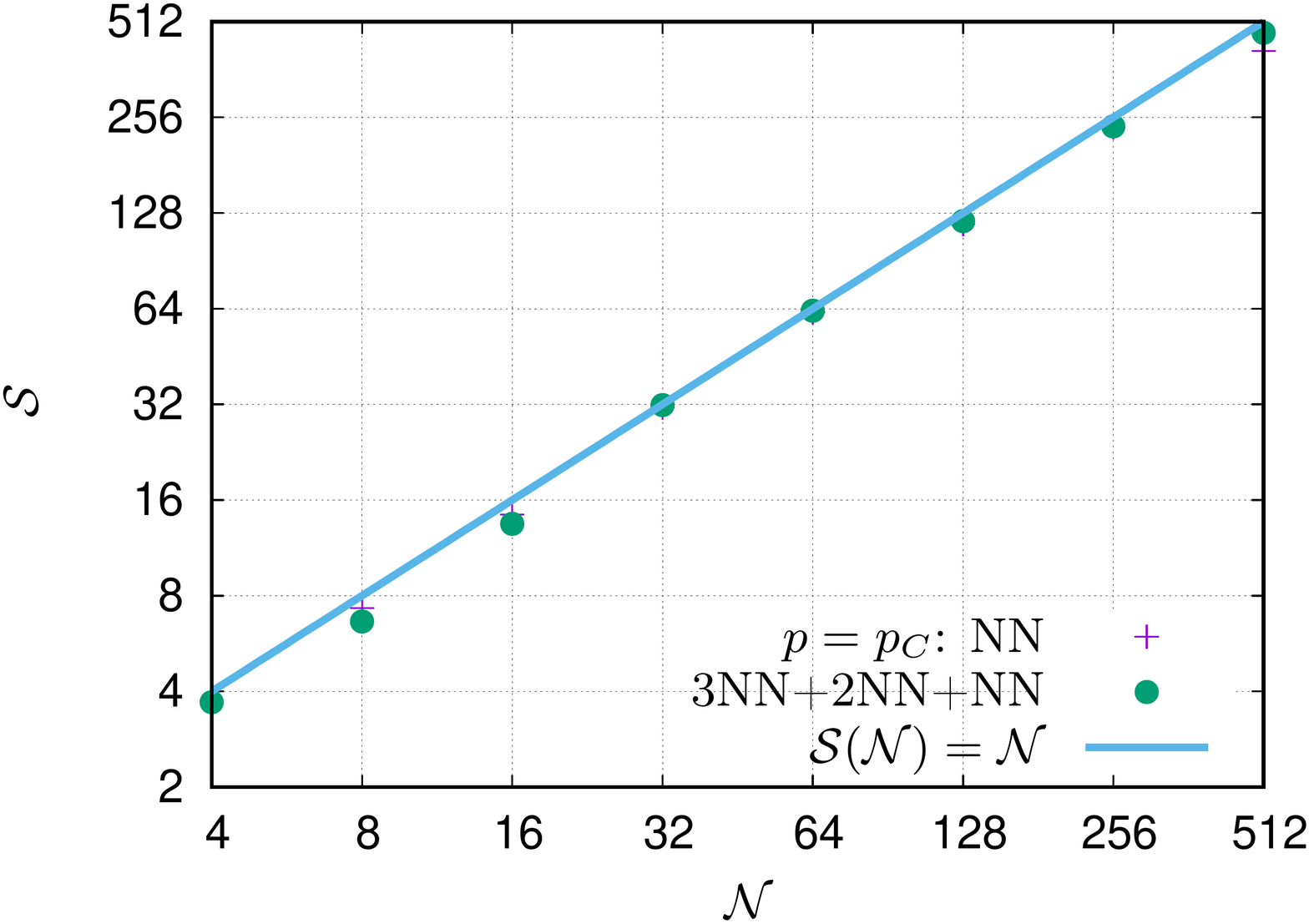}
\caption{\label{fig-S} (Colour online).
	Speed-up $\mathcal{S}$ evaluation for sites labeling at $p=p_C$ for NN and 3NN+2NN+NN neighbourhoods and various number of processors $\mathcal{N}$.
	The values of $\tau(L^d,1)$ are evaluated for $L=100$ and batch file (Listing~\ref{lst:batch}) parameters `\texttt{-N 1}' (single node) and `\texttt{--ntasks-per-node=1}' (single core), while for $\tau(L^d,\mathcal{N})$ evaluation we set `\texttt{-N 4}' (and subsequently `\texttt{-N 8}', `\texttt{-N 16}', `\texttt{-N 32}', `\texttt{-N 64}', `\texttt{-N 128}', `\texttt{-N 256}', `\texttt{-N 512}') and again for a signle core (`\texttt{--ntasks-per-node=1}'). 
The results are averaged over $R=10^4$ simulations.}
\end{figure}

Fig.~\ref{fig-S} demonstrates minimal parallel overhead observed during computations, due to negligible communication costs (see Sec.~\ref{sec-parallelism}).


\subsection{\label{sec-threshold}Percolation thresholds}

\begin{figure*}
\psfrag{L=}[][c]{{\small $L=$}}
\psfrag{p}{$p$}
\psfrag{W}[][c]{$W(p;L)$}
\psfrag{L}[][c]{$\lambda(p)$}
	\begin{subfigure}[b]{0.329\textwidth}
\caption{NN\label{fig-xNN}}
\includegraphics[width=\textwidth]{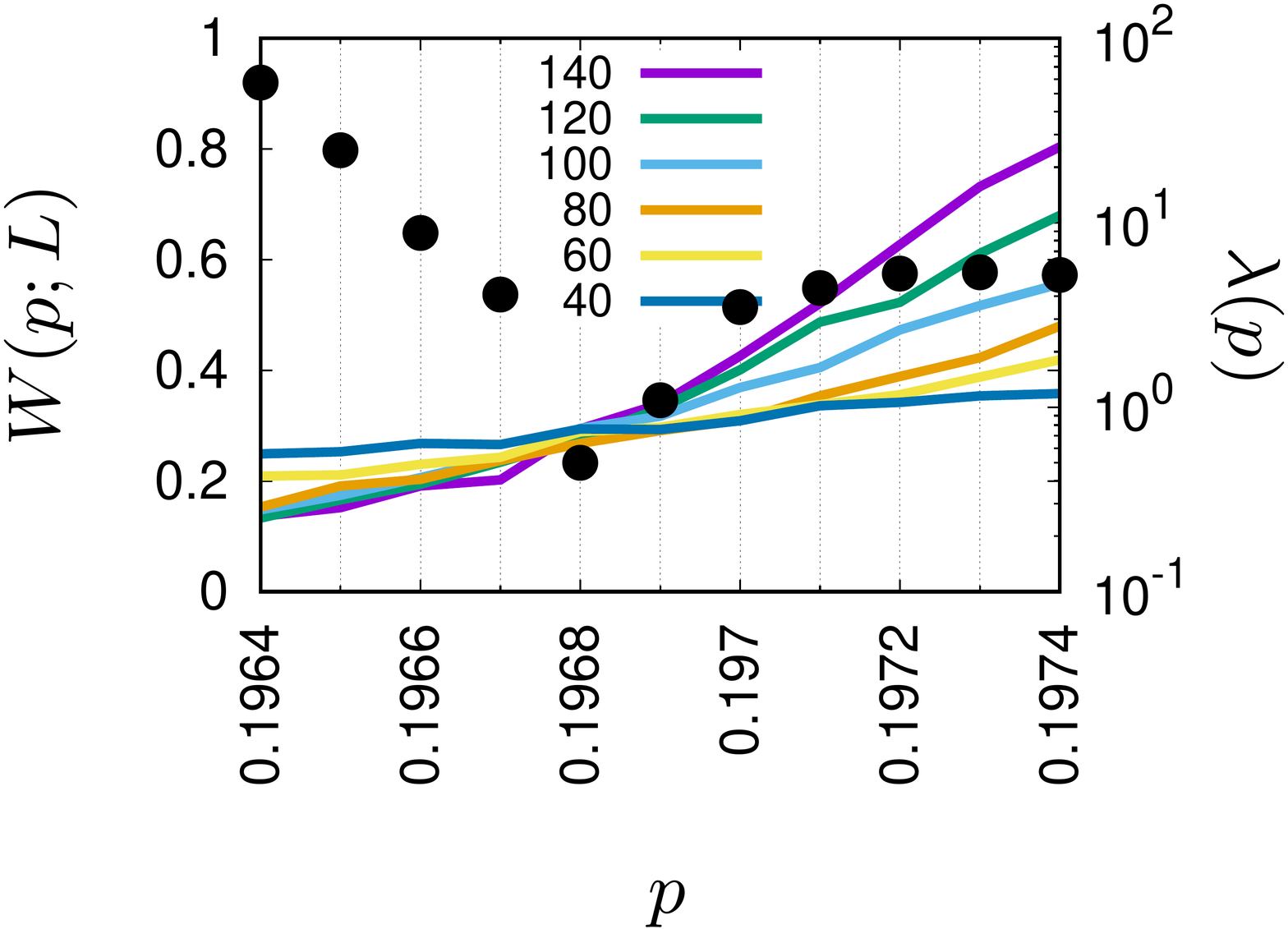}
\end{subfigure}\\
\begin{subfigure}[b]{0.329\textwidth}\caption{2NN\label{fig-x2NN}}{\includegraphics[width=\textwidth]{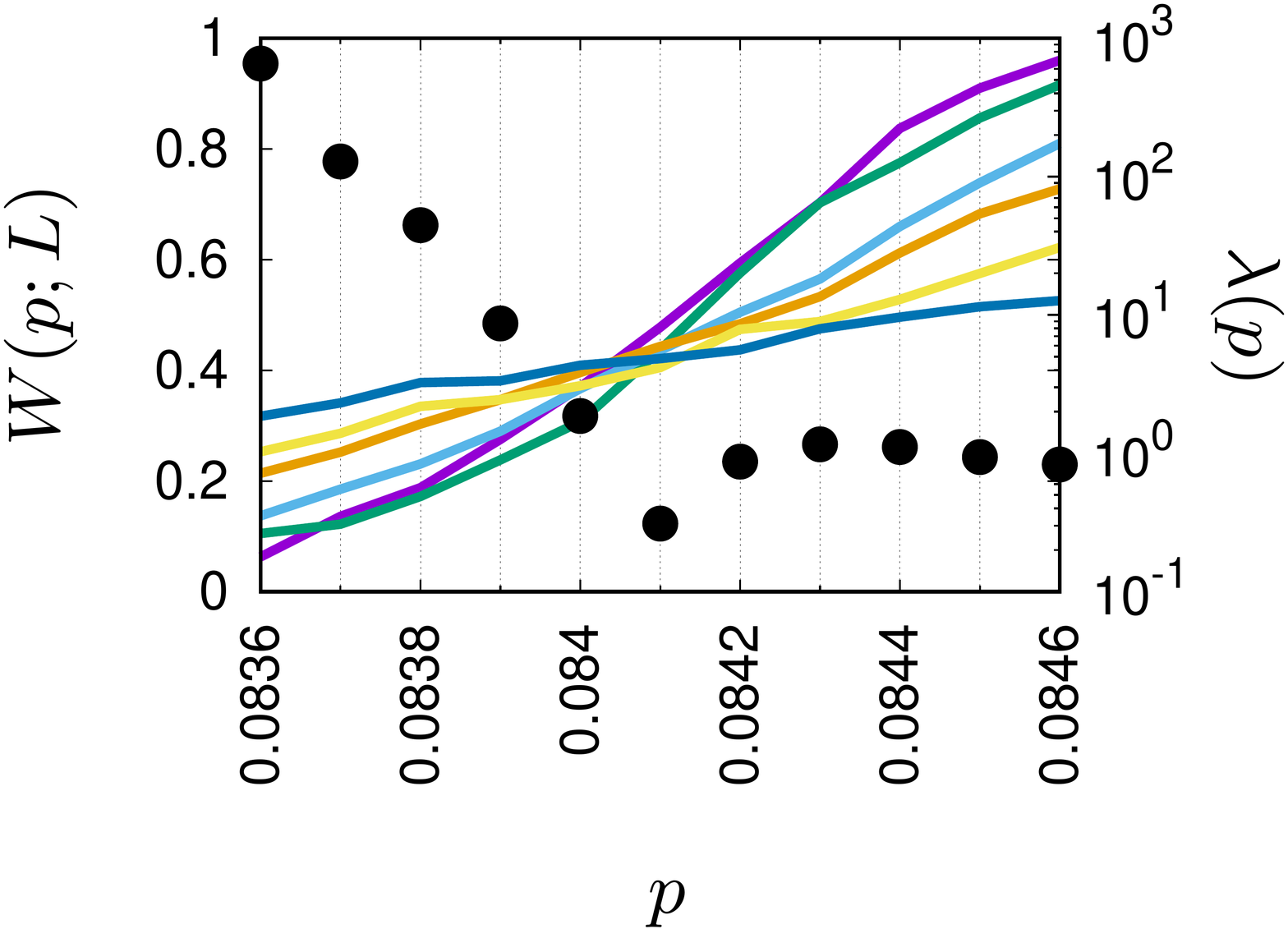}}\end{subfigure}
\hfill
\begin{subfigure}[b]{0.329\textwidth}\caption{2NN+NN\label{fig-x2NN_NN}}{\includegraphics[width=\textwidth]{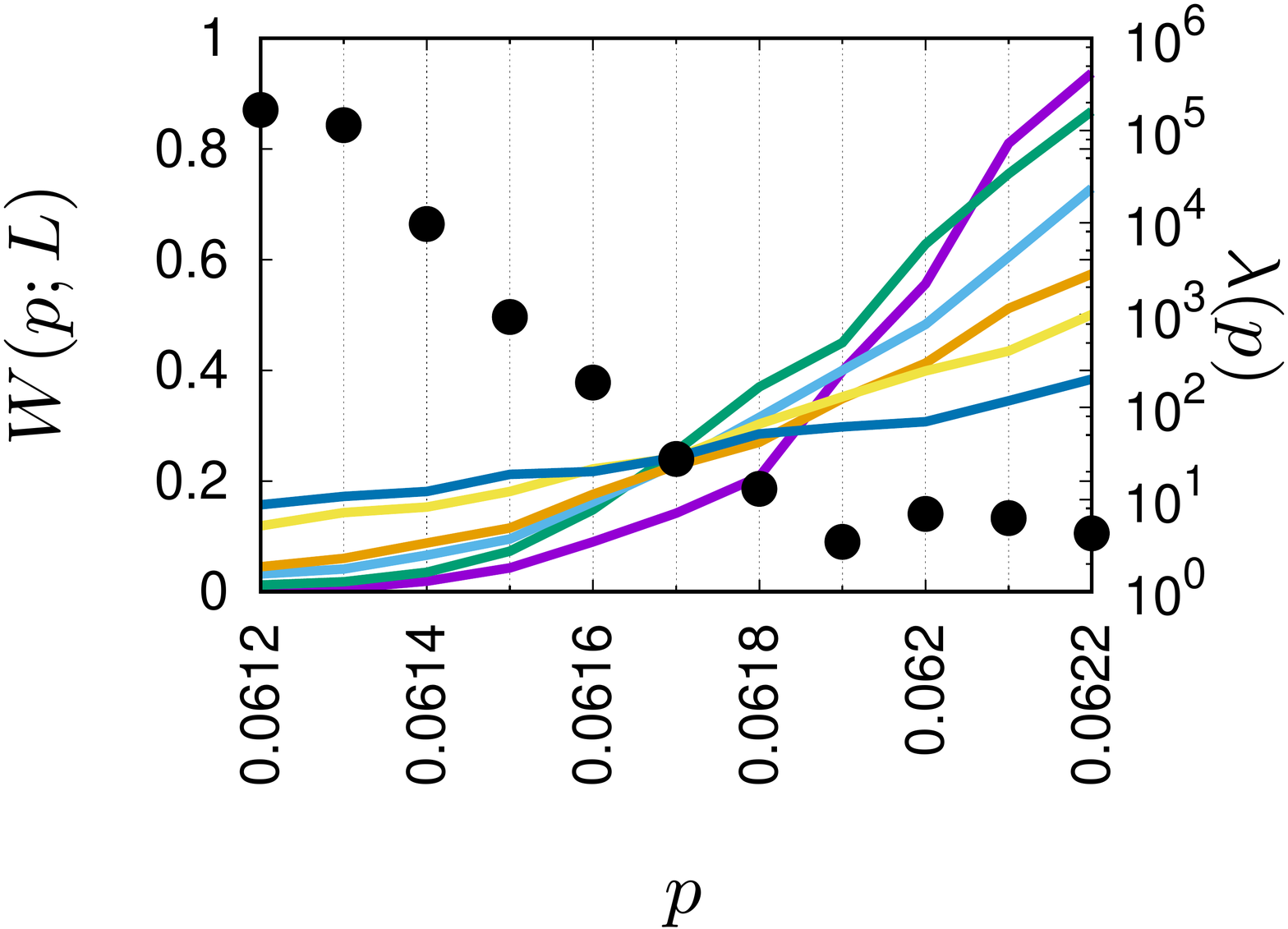}}\end{subfigure}
\hfill
\begin{subfigure}[b]{0.329\textwidth}\caption{3NN\label{fig-x3NN}}{{\includegraphics[width=\textwidth]{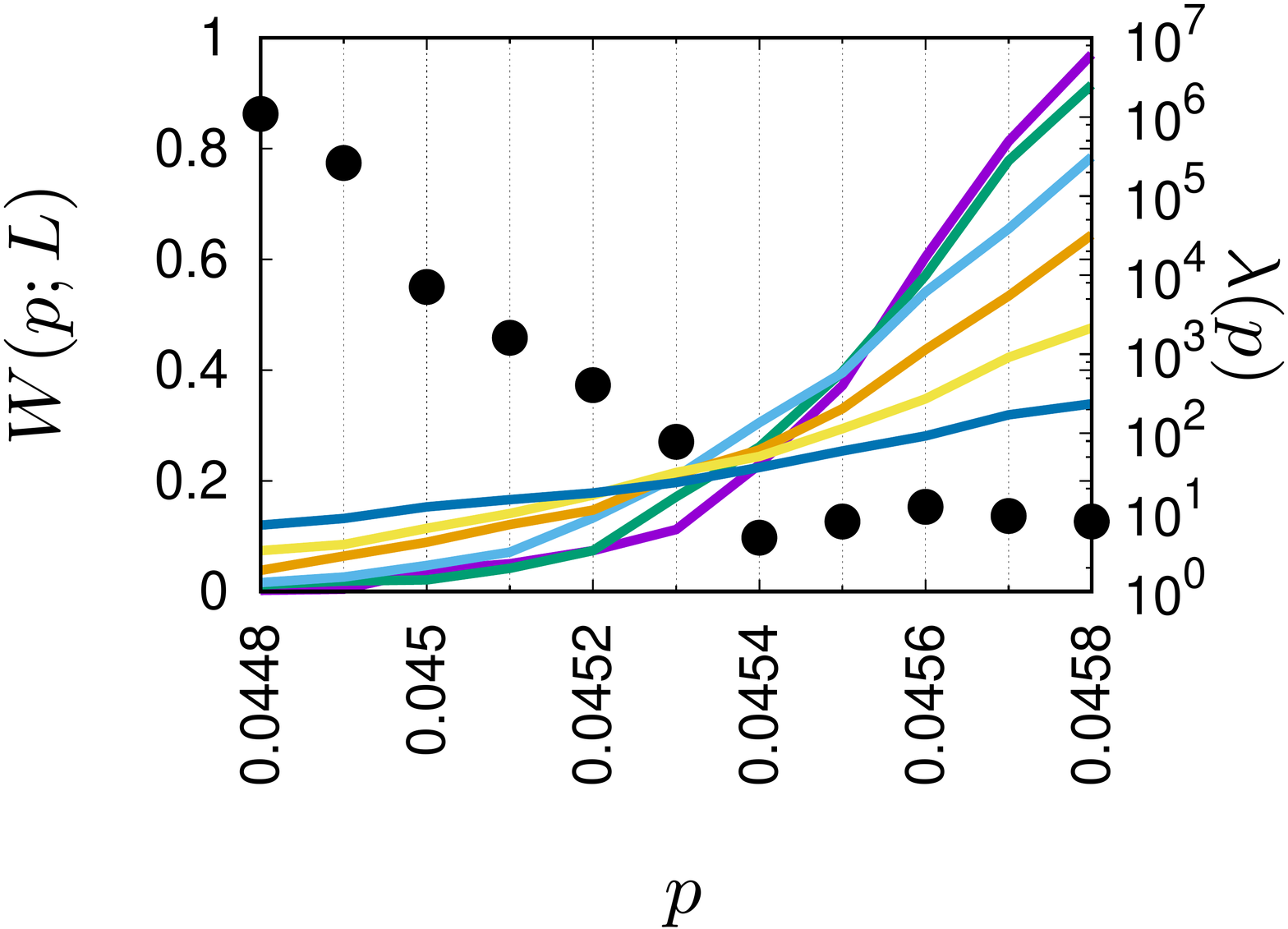}}}\end{subfigure}\\
\begin{subfigure}[b]{0.329\textwidth}\caption{3NN+NN\label{fig-x3NN_NN}}{{\includegraphics[width=\textwidth]{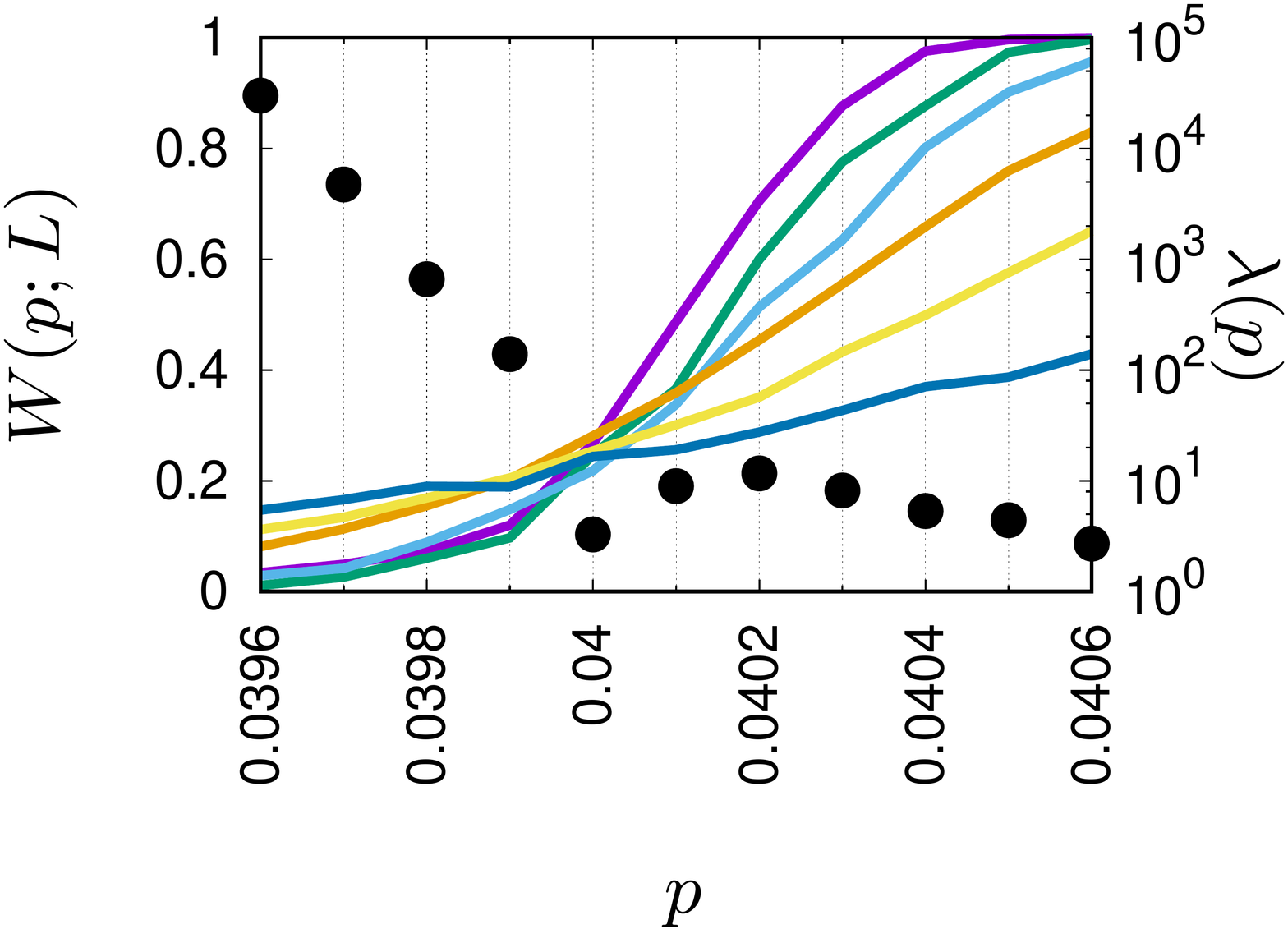}}}\end{subfigure}
\hfill
\begin{subfigure}[b]{0.329\textwidth}\caption{3NN+2NN\label{fig-x3NN_2NN}}{{\includegraphics[width=\textwidth]{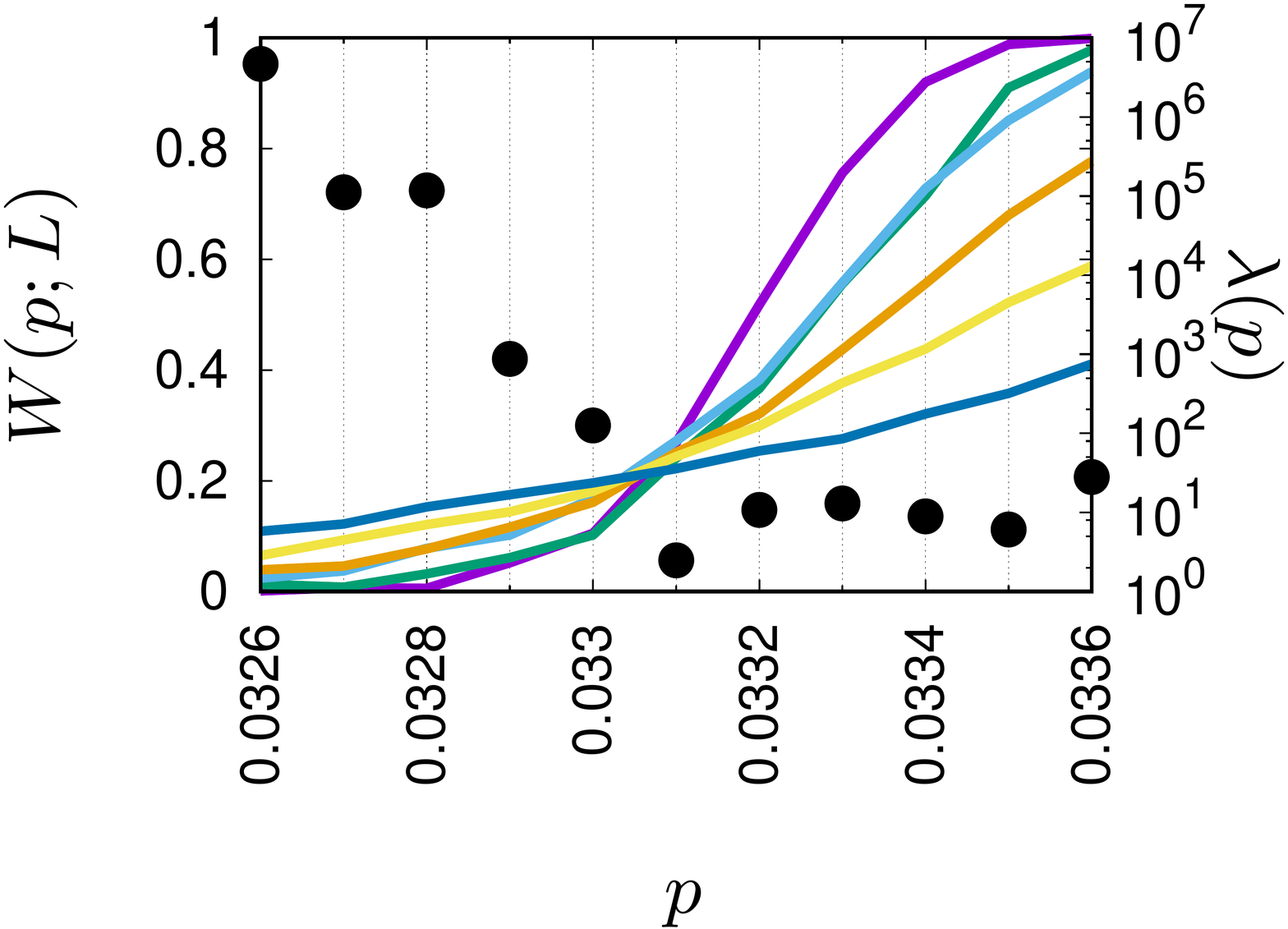}}}\end{subfigure}
\hfill
\begin{subfigure}[b]{0.329\textwidth}\caption{3NN+2NN+NN\label{fig-x3NN_2NN_NN}}{{\includegraphics[width=\textwidth]{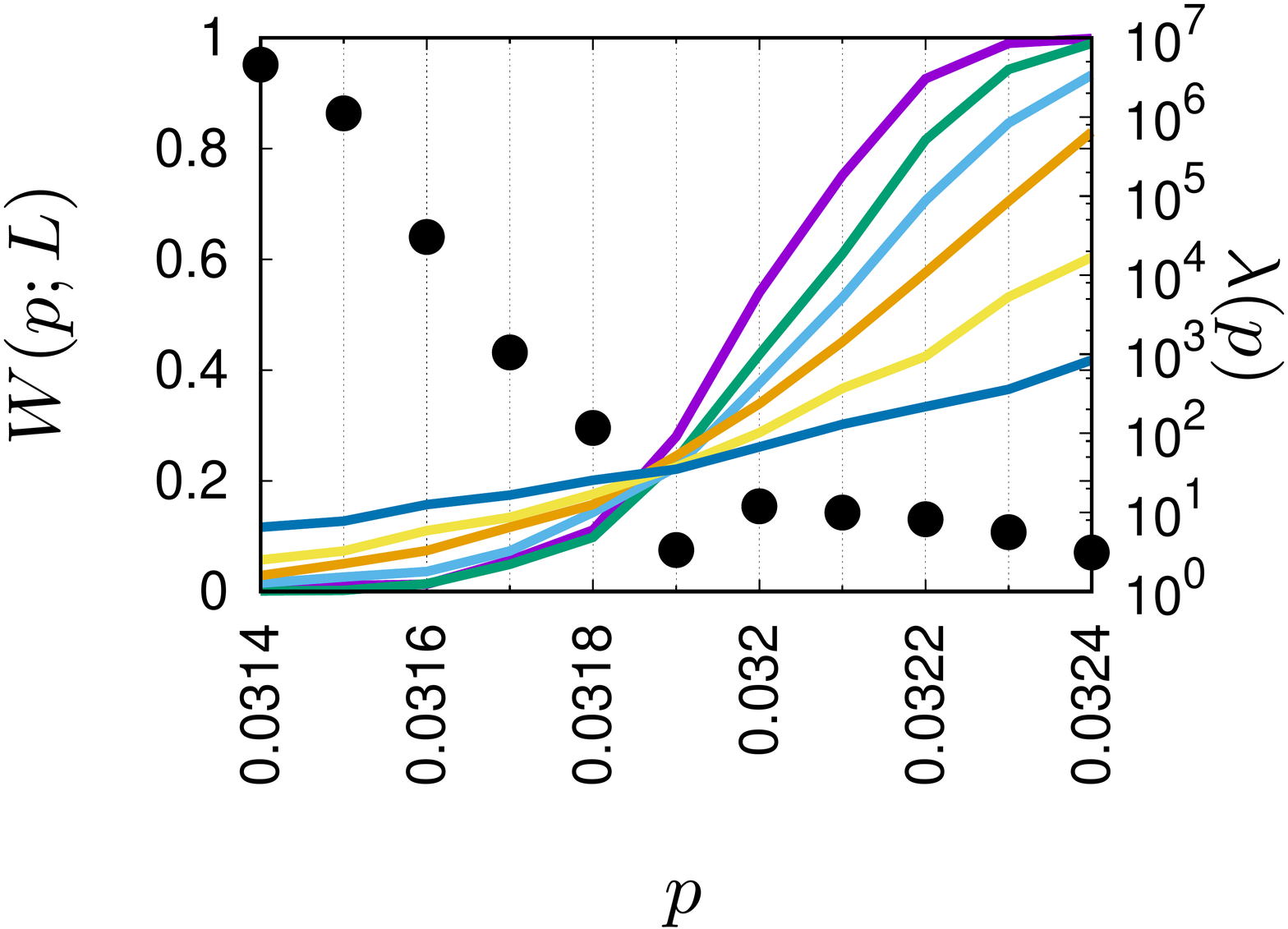}}}\end{subfigure}
\caption{\label{fig-xW} (Colour online). Wrapping probability $W(p;L)$ (lines, left axis) and the value $\lambda(p)$ (symbols, right axis) vs. occupation probability $p$.
The results are averaged over $R=10^5$ simulations.}
\end{figure*}

In Fig.~\ref{fig-xW} we plot wrapping probabilities $W(p;L)$ vs. occupation probability $p$ for various systems sizes $L$ ($40\le L\le 140)$ and various complex neighbourhoods (which contain various neighbours from NN to 3NN+2NN+NN).
In the same figure we also present the dependencies $\lambda(p)$ in semi-logarithmic scale.
The local minimum of $\lambda(p)$ near the interception of $W(p;L)$ for various $L$ indicates the percolation threshold $p_C$ obtained with Bastas et al. algorithm.

As we can see in Fig.~\ref{fig-xW} the true value of $p_C$ is hidden in the interval of the length of $2\Delta p$, where $\Delta p=10^{-4}$ is the scanning step of occupation probability $p$.
With believe that true value of $p_C$ is homogeneously distributed in this interval we can estimate the uncertainty of the percolation threshold $u(p_C)=2\Delta p/\sqrt{3}\approx 0.00023$.

\begin{table}[!htbp]
	\caption{\label{tab-PT} The values of percolation thresholds $p_C$ for SC lattice in four dimensional space ($d=4$) and for various neighbourhoods as deduced from Fig.~\ref{fig-xW}.}
\begin{ruledtabular}
\begin{tabular}{lrl}
 neighbourhood&$z$& $p_C$         \\ \hline
 NN          &  8& $0.19680(23)$ \\
 2NN         & 24& $0.08410(23)$ \\
 2NN+NN      & 32& $0.06190(23)$ \\
 3NN         & 32& $0.04540(23)$ \\
 3NN+NN      & 40& $0.04000(23)$ \\
 3NN+2NN     & 58& $0.03310(23)$ \\
 3NN+2NN+NN  & 64& $0.03190(23)$ \\
\end{tabular}
\end{ruledtabular}
\end{table}

The estimated values of $p_C$ together with its uncertainties are collected in Tab.~\ref{tab-PT}.
The table shows also the coordination numbers $z$ of sites for every considered complex neighbourhoods ranging from NN to 3NN+2NN+NN.

{Please note, that system with 2NN neighbours corresponds to 4-$d$ FCC lattice.
The obtained threshold $p_C(\text{2NN})\approx 0.08410(23)$ agrees within the error bars with earlier estimation $p_C(\text{FCC})\approx 0.0843(3)$  mentioned in Tab.}~\ref{tab-PT-ref}.

\section{\label{sec-conslusions}Conclusions}

In this paper the memory virtualization for the Hoshen--Kopelman~\cite{Hoshen1976a} is presented.
Due to minimal and constant cost of communication between processes the perfect speed-up ($\mathcal{S}(\mathcal{N})\approx\mathcal{N}$) is observed.

The achieved speed-up allows for computation---in reasonable time---the wrapping probabilities $W(p;L)$ (Eq.~\eqref{eq-W}) up to linear size of $L=140$ in $d=4$ dimensional space, i.e. for systems containg $3.8416\times 10^{8}$ sites {realized $10^4$ times}.

The finite-size scaling technique (see Eq.~\eqref{eq-scaling} and Sec.~\ref{sec-approach}) combined with Bastas et al. technique \cite{PhysRevE.84.066112,Bastas2014,Malarz2015} allows for estimation of percolation thresholds for simple cubic lattice in $d=4$ and for neighbourhoods ranging from NN to 3NN+2NN+NN with the accuracy $u(p_C)=23\times 10^{-5}$.

The estimated values of $p_C$ together with its uncertainties are collected in Tab.~\ref{tab-PT}.
Our results enriches earlier studies regarding percolation thresholds for complex neighbourhoods on square~\cite{Galam2005a,*Galam2005b,*Majewski2007} or three-dimensional SC~\cite{Kurzawski2012,Malarz2015} lattices.

\begin{acknowledgments}
This work was financed by PL-Grid infrastructure and partially supported by AGH-UST statutory tasks No. 11.11.220.01/2 within subsidy of the Ministry of Science and Higher Education.
\end{acknowledgments}

\appendix

\section{Program description}

The program (Listing~\ref{lst:main}) was written in Fortran 95~\cite{Fortran95} and compiled using Intel Fortran compiler ifort. It follows purely procedural paradigm while utilising some array-manipulation features.

The programs parameters are read from command line. All of them are integers and should be simply written one after another, as all of them are mandatory. They are the following:
\begin{itemize}
  \item \varname{L} --- linear size of the problem,
  \item \varname{p\_min} --- minimal occupation probability to be checked,
  \item \varname{p\_max} --- maximal occupation probability to be checked,
  \item \varname{p\_step} --- loop step over probabilities range,
  \item \varname{N\_run} --- number of simulations for each value of $p$.
\end{itemize}

All the parameters and dynamically allocated memory is the same through all the program, which loops through over the examined range of probability from \varname{p\_min} to \varname{p\_max} with step \varname{p\_step}. For any given $p$, all processes run subsequent tasks in parallel to each other, only sending results to the main process (through \varname{MPI\_sum}) after finishing all the tasks for the given value of $p$.

Sites are given labels according to Hoshen--Kopelman algorithm. There are two kinds of special labels: \varname{FILL\_LABEL}, having the lowest possible value, and \varname{EMPTY\_LABEL}, having the highest one. Each new casual label is given a successively increasing number. Thanks to that, reclassifying requires no filling checking (although aliasing still does) and checking for percolation is a simple comparison of labels after reclassification: any label with value less than the last label of the first slice is percolating (\procname{has\_percolation} function).

Hyper-cubic space is virtualised (see Sec.~\ref{sec-implementation}) with merged filling and labelling (conditional at l.~173). Then if a new cluster is made (function \procname{is\_new\_cluster}), the site is given a label represented by a successive integer number. If not, cluster surrounding the site are merged (subroutine \procname{merge\_cluster}). At the end of the buffer (l.~202-209), the last line is copied before the starting line and the process starts again until the virtual depth of \varname{L}. The first slice is additionally stored separately (l.~191-199).

The program contains basic time measurement.

\section{Source files}

The program uses main program (Listing~\ref{lst:main}), makefile (Listing~\ref{lst:makefile}) and batch file (Listing~\ref{lst:batch}).

Makefile is a standard single-target building tool. It provides two commands: `build' (or `all') and `clean'.
Four parameters can be changed:
\begin{itemize}
  \item FC --- Fortran compiler to be used,
  \item PAR\_FC --- Fortran parallel (MPI) compiler to be used,
  \item RM --- cleaning operation,
  \item SRC --- source file.
\end{itemize}

Batch file runs the executable as a new task on cluster. It contains both program's run-time parameters and task parameters. Task parameters are the following:
\begin{itemize}
	\item \texttt{-J <name>} --- name of the task,
	\item \texttt{-N <number>} --- number of nodes to run the task on,
	\item \texttt{--ntasks-per-node=<number>} --- number of tasks per node, optimally the same as the number of cores on the given architecture,
	\item \texttt{--time=<HH:MM:SS>} --- time limit for the task,
	\item \texttt{-A <grant>} --- name of the computational grant,
	\item \texttt{--output=<file>} --- file to which the output will be redirected,
	\item \texttt{--error=<file>} --- file to which the error information will be redirected.
\end{itemize}

\bibliography{percolation,MPI,km}

\definecolor{mygray}{rgb}{0.92,0.97,0.92}
\lstset{backgroundcolor=\color{mygray},commentstyle=\itshape\color{blue},stringstyle=\color{teal}}

\begin{widetext}
\begin{lstlisting}[language={[95]Fortran},frame=single,numbers=left,numberstyle=\tiny,basicstyle=\footnotesize,stepnumber=1,breaklines=true,caption={Fortan95 code allowing for direct reproduction of the results presented in Figs.~\ref{fig-W} and \ref{fig-xW}.},label=lst:main]
program pi_4d_2n3n4n
  use ifport, only: rand
  use mpi
  implicit none

  ! time measurement
  real :: start, finish
  
  ! random seed
  integer :: seed

  ! loop counters
  integer :: i_main, i, j, k, m, hb
  integer :: irun, N_run, maxlabel

  ! maximum L so that the whole hypercube fits into the buffer,
  ! determines buffer's size as (MAX_L+4)**4
  integer, parameter :: MAX_L = 110

  ! minimum depth of the buffer
  ! the lower, the flatter it can get (and more copying is involved)
  ! depth is the first dimension (i) so the copying cost should be minimal
  integer, parameter :: MIN_DEPTH = 10

  ! hypercube's edge's length (and length with reserved space)
  ! experimental limit is L = 418 (L+4 == 422)
  integer :: L, L4

  ! probability of filling a site
  real ::  p, p_min, p_max, p_step

  ! percentage of systems with a percolating cluster
  double precision :: perc_local, perc

  ! label for each site
  integer, dimension(:,:,:,:), allocatable :: label

  ! label reduction (aliasing) table
  integer, dimension(:), allocatable :: iclass

  ! label for filled sites, has to be smaller than any other label
  integer :: FILL_LABEL

  ! label for empty sites, has to be greater than any other label
  integer :: EMPTY_LABEL

  ! communication
  integer :: world_comm, nproc, rank, root_rank, ierror

  ! string buffer
  character(len=20) :: str_buffer

  !---------------------------------------------------------------------------


  ! start communication

  world_comm = Mpi_Comm_World
  call Mpi_Init(ierror)
  call Mpi_Comm_Size(world_comm, nproc, ierror)
  call Mpi_Comm_Rank(world_comm, rank , ierror)
  root_rank = 0

  !---------------------------------------------------------------------------
  ! start random engine

  call system_clock(seed)  ! any random seed, basically
  call srand(seed)

  !---------------------------------------------------------------------------
  ! read options

  if(rank == root_rank) then
    call cpu_time(start)
  end if

  !---------------------------------------------------------------------------
  ! read parameters

  if(command_argument_count() < 5) then
    if(rank == root_rank) then
      print *, 'Too few arguments, expected 5: L, p_min, p_max, p_step, N_run'
    end if
    call abort
  end if

  str_buffer = ""
  call get_command_argument(1, str_buffer)
  read(str_buffer, fmt="(I)") L
  str_buffer = ""
  call get_command_argument(2, str_buffer)
  read(str_buffer, fmt="(F)") p_min
  str_buffer = ""
  call get_command_argument(3, str_buffer)
  read(str_buffer, fmt="(F)") p_max
  str_buffer = ""
  call get_command_argument(4, str_buffer)
  read(str_buffer, fmt="(F)") p_step
  str_buffer = ""
  call get_command_argument(5, str_buffer)
  read(str_buffer, fmt="(I)") N_run

  L4 = L**4

  if(rank == root_rank) then
    print *, '# 4D: 2N3N4N'
    print *, '#   L   N_run   L4'
    print *, '#', L, N_run, L4
  end if

  !---------------------------------------------------------------------------
  ! read parameters
  ! allocate buffer, cluster classes etc.

  associate(depth => space_depth(L+4, L+4, L+4, L+4))
    if(depth /= L+4) then
      ! buffering can even give a minor speedup
      if(depth > MIN_DEPTH) then
        if(rank == root_rank) then
          print *, '# Big L value. Buffering with size: ', depth
        end if
      else
        if(rank == root_rank) then
          print *, '# Too big L value. Abort'
        end if
        call abort
      end if
    end if

    allocate(label(depth, L+4, L+4, L+4))
    allocate(iclass(L4/2+2))
    FILL_LABEL  = lbound(iclass, 1)
    EMPTY_LABEL = ubound(iclass, 1)
  end associate

  !---------------------------------------------------------------------------
  ! main calculations part

  ! all processes share the same probability...
  do p = p_min, p_max, p_step

    perc_local = 0.0d0

    ! ...but each process calculates a separate simulation
    do irun = rank+1, N_run, nproc
      call initialize(maxlabel, label, iclass)

      ! over the edge with buffer depth as step
      do i_main = 0, L-1, size(label,1)-4
        hb = min(ubound(label,1)-2, L-1 - i_main + lbound(label,1)+2)

        do i = lbound(label,1)+2, hb
          do j = lbound(label,2)+2, ubound(label,2)-2
            do k = lbound(label,3)+2, ubound(label,3)-2
              do m = lbound(label,4)+2, ubound(label,4)-2
                ! labeling clusters

                ! fill randomly + check whether filled or not == just rand
                if(rand() < p) then
                  if(is_new_cluster(label, i, j, k, m)) then
                    maxlabel = maxlabel+1
                    label(i,j,k,m) = maxlabel
                  else
                    call merge_clusters(label, i, j, k, m, iclass)
                  end if
                else
                  label(i,j,k,m) = EMPTY_LABEL
                end if
              end do
            end do
          end do
        end do

        if(i_main < (L+2) - (size(label,1)-4)) then
          ! 3D space first in depth
          ! saved in unused last space as it will be used later
          if(i_main == 0) then
            do j = lbound(label,2)+2, ubound(label,2)-2
              do k = lbound(label,3)+2, ubound(label,3)-2
                do m = lbound(label,4)+2, ubound(label,4)-2
                  label(ubound(label,1), j, k, m) = label(lbound(label,1)+2, j, k, m)
                end do 
              end do 
            end do
          end if

          ! wrapping buffer around
          do j = lbound(label,2)+2, ubound(label,2)-2
            do k = lbound(label,3)+2, ubound(label,3)-2
              do m = lbound(label,4)+2, ubound(label,4)-2
                label(lbound(label,1)+0, j, k, m) = label(hb-1, j, k, m)
                label(lbound(label,1)+1, j, k, m) = label(hb-0, j, k, m)
              end do 
            end do 
          end do
        end if
      end do

      ! checking if a cluster is spanning whole system

      ! first 3D space in depth is stored in unused buffer's part
      maxlabel = max_source_label(label, ubound(label,1))

      if(has_percolation(label, hb, maxlabel)) then
        perc_local = perc_local + 1.0d0/(1.0d0*N_run)
      end if
    end do

  !---------------------------------------------------------------------------
  ! get stats from all processes into root process

    call Mpi_Reduce(perc_local, perc, 1, Mpi_Double_Precision, Mpi_Sum, &
                    root_rank, world_comm, ierror)

    ! combine stats
    if(rank == root_rank) then
      print '(F10.6,1X,F7.3)', p, perc
    end if
  end do

  !---------------------------------------------------------------------------
  ! deallocate buffer, cluster classes etc.

  deallocate(label)
  deallocate(iclass)

  !---------------------------------------------------------------------------
  ! show time information

  if(rank == root_rank) then
    call cpu_time(finish)
    print '("# Time = ",f20.3," seconds.")', finish-start
  end if

  !---------------------------------------------------------------------------
  ! finalize communication
  call Mpi_Finalize(ierror)


contains
  !---------------------------------------------------------------------------
  ! calculate space depth
  integer pure function space_depth(N, M, K, L)
    integer, intent(in) :: N, M, K, L

    space_depth = min((MAX_L+4)**4 / (M*K*L) - 4, N)
  end function space_depth

  !---------------------------------------------------------------------------
  ! initialize perlocation to zero
  subroutine initialize(maxlabel, label, iclass)
    integer, intent(out) :: maxlabel
    integer, intent(out), dimension(:,:,:,:) :: label
    integer, intent(out), dimension(:) :: iclass

    maxlabel = FILL_LABEL + 1
    label(:,:,:,:) = EMPTY_LABEL

    do i = lbound(iclass, 1), ubound(iclass, 1)
      iclass(i) = i
    end do
  end subroutine

  !---------------------------------------------------------------------------
  ! whether the given site makes a new cluster
  logical pure function is_new_cluster(label, i, j, k, m)
    integer, intent(in), dimension(:,:,:,:) :: label
    integer, intent(in) :: i, j, k, m

    is_new_cluster = label(i-1,j  ,k  ,m  ) == EMPTY_LABEL .and. & 
                     label(i  ,j-1,k  ,m  ) == EMPTY_LABEL .and. &
                     label(i  ,j  ,k-1,m  ) == EMPTY_LABEL .and. &
                     label(i  ,j  ,k  ,m-1) == EMPTY_LABEL .and. &

                     label(i-1,j-1,k  ,m  ) == EMPTY_LABEL .and. &
                     label(i-1,j+1,k  ,m  ) == EMPTY_LABEL .and. &
                     label(i-1,j  ,k-1,m  ) == EMPTY_LABEL .and. &
                     label(i-1,j  ,k+1,m  ) == EMPTY_LABEL .and. &
                     label(i-1,j  ,k  ,m-1) == EMPTY_LABEL .and. &
                     label(i-1,j  ,k  ,m+1) == EMPTY_LABEL .and. &
                     label(i  ,j-1,k-1,m  ) == EMPTY_LABEL .and. &
                     label(i  ,j-1,k+1,m  ) == EMPTY_LABEL .and. &
                     label(i  ,j-1,k  ,m-1) == EMPTY_LABEL .and. &
                     label(i  ,j-1,k  ,m+1) == EMPTY_LABEL .and. &
                     label(i  ,j  ,k-1,m-1) == EMPTY_LABEL .and. &
                     label(i  ,j  ,k-1,m+1) == EMPTY_LABEL .and. &

                     label(i  ,j-1,k-1,m-1) == EMPTY_LABEL .and. & 
                     label(i  ,j-1,k+1,m-1) == EMPTY_LABEL .and. &
                     label(i  ,j-1,k-1,m+1) == EMPTY_LABEL .and. &
                     label(i  ,j-1,k+1,m+1) == EMPTY_LABEL .and. &
                     label(i-1,j  ,k-1,m-1) == EMPTY_LABEL .and. &
                     label(i-1,j  ,k+1,m-1) == EMPTY_LABEL .and. &
                     label(i-1,j  ,k-1,m+1) == EMPTY_LABEL .and. &
                     label(i-1,j  ,k+1,m+1) == EMPTY_LABEL .and. &
                     label(i-1,j-1,k  ,m-1) == EMPTY_LABEL .and. &
                     label(i-1,j+1,k  ,m-1) == EMPTY_LABEL .and. &
                     label(i-1,j-1,k  ,m+1) == EMPTY_LABEL .and. &
                     label(i-1,j+1,k  ,m+1) == EMPTY_LABEL .and. &
                     label(i-1,j-1,k-1,m  ) == EMPTY_LABEL .and. &
                     label(i-1,j+1,k-1,m  ) == EMPTY_LABEL .and. &
                     label(i-1,j-1,k+1,m  ) == EMPTY_LABEL .and. &
                     label(i-1,j+1,k+1,m  ) == EMPTY_LABEL
  end function is_new_cluster

  !---------------------------------------------------------------------------
  ! merge clusters connecting at the given point
  subroutine merge_clusters(label, i, j, k, m, iclass)
    integer, intent(inout), dimension(:,:,:,:) :: label
    integer, intent(in) :: i, j, k, m
    integer, intent(inout), dimension(:) :: iclass

    ! reclassifing neighbours
    ! 2N:
    call reclassify(label(i-1,j  ,k  ,m  ), iclass)
    call reclassify(label(i  ,j-1,k  ,m  ), iclass)
    call reclassify(label(i  ,j  ,k-1,m  ), iclass)
    call reclassify(label(i  ,j  ,k  ,m-1), iclass)

    ! 3N:
    call reclassify(label(i-1,j-1,k  ,m  ), iclass)
    call reclassify(label(i-1,j+1,k  ,m  ), iclass)
    call reclassify(label(i-1,j  ,k-1,m  ), iclass)
    call reclassify(label(i-1,j  ,k+1,m  ), iclass)
    call reclassify(label(i-1,j  ,k  ,m-1), iclass)
    call reclassify(label(i-1,j  ,k  ,m+1), iclass)

    call reclassify(label(i  ,j-1,k-1,m  ), iclass)
    call reclassify(label(i  ,j-1,k+1,m  ), iclass)
    call reclassify(label(i  ,j-1,k  ,m-1), iclass)
    call reclassify(label(i  ,j-1,k  ,m+1), iclass)

    call reclassify(label(i  ,j  ,k-1,m-1), iclass)
    call reclassify(label(i  ,j  ,k-1,m+1), iclass)

    ! 4N:
    call reclassify(label(i-1,j-1,k-1,m  ), iclass)
    call reclassify(label(i-1,j+1,k-1,m  ), iclass)
    call reclassify(label(i-1,j-1,k+1,m  ), iclass)
    call reclassify(label(i-1,j+1,k+1,m  ), iclass)

    call reclassify(label(i-1,j  ,k-1,m-1), iclass)
    call reclassify(label(i-1,j  ,k+1,m-1), iclass)
    call reclassify(label(i-1,j  ,k-1,m+1), iclass)
    call reclassify(label(i-1,j  ,k+1,m+1), iclass)

    call reclassify(label(i-1,j-1,k  ,m-1), iclass)
    call reclassify(label(i-1,j+1,k  ,m-1), iclass)
    call reclassify(label(i-1,j-1,k  ,m+1), iclass)
    call reclassify(label(i-1,j+1,k  ,m+1), iclass)

    call reclassify(label(i  ,j-1,k-1,m-1), iclass)
    call reclassify(label(i  ,j-1,k+1,m-1), iclass)
    call reclassify(label(i  ,j-1,k-1,m+1), iclass)
    call reclassify(label(i  ,j-1,k+1,m+1), iclass)

    ! good label according to rule:
    !   a <= b  =>  iclass(a) <= iclass(b)
    label(i,j,k,m) = min(     &
      label(i-1,j  ,k  ,m  ), &
      label(i  ,j-1,k  ,m  ), &
      label(i  ,j  ,k-1,m  ), &
      label(i  ,j  ,k  ,m-1), &

      label(i-1,j-1,k  ,m  ), &
      label(i-1,j+1,k  ,m  ), &
      label(i-1,j  ,k-1,m  ), &
      label(i-1,j  ,k+1,m  ), &
      label(i-1,j  ,k  ,m-1), &
      label(i-1,j  ,k  ,m+1), &
      label(i  ,j-1,k-1,m  ), &
      label(i  ,j-1,k+1,m  ), &
      label(i  ,j-1,k  ,m-1), &
      label(i  ,j-1,k  ,m+1), &
      label(i  ,j  ,k-1,m-1), &
      label(i  ,j  ,k-1,m+1), &

      label(i  ,j-1,k-1,m-1), &
      label(i  ,j-1,k+1,m-1), &
      label(i  ,j-1,k-1,m+1), &
      label(i  ,j-1,k+1,m+1), &
      label(i-1,j  ,k-1,m-1), &
      label(i-1,j  ,k+1,m-1), &
      label(i-1,j  ,k-1,m+1), &
      label(i-1,j  ,k+1,m+1), &
      label(i-1,j-1,k  ,m-1), &
      label(i-1,j+1,k  ,m-1), &
      label(i-1,j-1,k  ,m+1), &
      label(i-1,j+1,k  ,m+1), &
      label(i-1,j-1,k-1,m  ), &
      label(i-1,j+1,k-1,m  ), &
      label(i-1,j-1,k+1,m  ), &
      label(i-1,j+1,k+1,m  ))

    ! make the label common as iclass
    call make_alias(label, i-1,j,k,m, i,j,k,m)
    call make_alias(label, i,j-1,k,m, i,j,k,m)
    call make_alias(label, i,j,k-1,m, i,j,k,m)
    call make_alias(label, i,j,k,m-1, i,j,k,m)

    call make_alias(label, i-1,j-1,k  ,m  , i,j,k,m)
    call make_alias(label, i-1,j+1,k  ,m  , i,j,k,m)
    call make_alias(label, i-1,j  ,k-1,m  , i,j,k,m)
    call make_alias(label, i-1,j  ,k+1,m  , i,j,k,m)
    call make_alias(label, i-1,j  ,k  ,m-1, i,j,k,m)
    call make_alias(label, i-1,j  ,k  ,m+1, i,j,k,m)
    call make_alias(label, i  ,j-1,k-1,m  , i,j,k,m)
    call make_alias(label, i  ,j-1,k+1,m  , i,j,k,m)
    call make_alias(label, i  ,j-1,k  ,m-1, i,j,k,m)
    call make_alias(label, i  ,j-1,k  ,m+1, i,j,k,m)
    call make_alias(label, i  ,j  ,k-1,m-1, i,j,k,m)
    call make_alias(label, i  ,j  ,k-1,m+1, i,j,k,m)

    call make_alias(label, i  ,j-1,k-1,m-1, i,j,k,m)
    call make_alias(label, i  ,j-1,k+1,m-1, i,j,k,m)
    call make_alias(label, i  ,j-1,k-1,m+1, i,j,k,m)
    call make_alias(label, i  ,j-1,k+1,m+1, i,j,k,m)
    call make_alias(label, i-1,j  ,k-1,m-1, i,j,k,m)
    call make_alias(label, i-1,j  ,k+1,m-1, i,j,k,m)
    call make_alias(label, i-1,j  ,k-1,m+1, i,j,k,m)
    call make_alias(label, i-1,j  ,k+1,m+1, i,j,k,m)
    call make_alias(label, i-1,j-1,k  ,m-1, i,j,k,m)
    call make_alias(label, i-1,j+1,k  ,m-1, i,j,k,m)
    call make_alias(label, i-1,j-1,k  ,m+1, i,j,k,m)
    call make_alias(label, i-1,j+1,k  ,m+1, i,j,k,m)
    call make_alias(label, i-1,j-1,k-1,m  , i,j,k,m)
    call make_alias(label, i-1,j+1,k-1,m  , i,j,k,m)
    call make_alias(label, i-1,j-1,k+1,m  , i,j,k,m)
    call make_alias(label, i-1,j+1,k+1,m  , i,j,k,m)
  end subroutine merge_clusters

  !---------------------------------------------------------------------------
  ! get maximal label for the 3D space first in depth
  integer pure function max_source_label(label, row)
    integer, intent(in), dimension(:,:,:,:) :: label
    integer, intent(in) :: row
    integer :: j, k, m

    max_source_label = 0
    do j = lbound(label,2)+2, ubound(label,2)-2
      do k = lbound(label,3)+2, ubound(label,3)-2
        do m = lbound(label,4)+2, ubound(label,4)-2
          if(label(row, j, k, m) /= EMPTY_LABEL) then
            max_source_label = max(max_source_label, &
                                   label(row, j, k, m))
          end if
        end do
      end do
    end do
  end function max_source_label

  !---------------------------------------------------------------------------
  ! whether there is a percolation or not
  logical pure function has_percolation(label, row, maxlabel)
    integer, intent(in), dimension(:,:,:,:) :: label
    integer, intent(in) :: row, maxlabel
    integer :: j, k, m

    has_percolation = .false.
    do j = lbound(label,2)+2, ubound(label,2)-2
      do k = lbound(label,3)+2, ubound(label,3)-2
        do m = lbound(label,4)+2, ubound(label,4)-2
          if(label(row, j, k, m) <= maxlabel) then
            has_percolation = .true.
            return
          end if
        end do
      end do
    end do
  end function has_percolation

  !---------------------------------------------------------------------------
  ! reduce site's label's aliasing
  ! (assign it its non-aliasing label)
  subroutine reclassify(ix, iclass)
    integer, intent(inout) :: ix
    integer, intent(in), dimension(:) :: iclass

    do while (iclass(ix) /= ix)
      ix = iclass(ix)
    end do
  end subroutine reclassify

  !---------------------------------------------------------------------------
  ! makes one site's cluster an alias for another one
  subroutine make_alias(label, i1, j1, k1, m1, i2, j2, k2, m2)
    integer, dimension(:,:,:,:), intent(inout) :: label
    integer, intent(in) :: i1,j1,k1,m1, i2,j2,k2,m2

    if(label(i1,j1,k1,m1) /= EMPTY_LABEL) then
      iclass(label(i1,j1,k1,m1)) = label(i2,j2,k2,m2)
    end if
  end subroutine


end program pi_4d_2n3n4n
\end{lstlisting}
\begin{lstlisting}[language={bash},frame=single,numbers=left,numberstyle=\tiny,basicstyle=\footnotesize,stepnumber=1,breaklines=true,caption={Makefile},label=lst:makefile]
RM=rm
FC=ifort
PAR_FC=mpiifort

FLAGS = -g -O3 -no-prec-div -fp-model fast=2 -xHost -ipo -warn all

SRC=pi_4d_2n3n4n.f90
OBJ=${SRC:.f90=.o}
EXE=${SRC:.f90=.exe}


.PHONY: all
all: build

.PHONY: build
build: $(EXE)


$(EXE): $(OBJ)
    $(PAR_FC) $(FLAGS) -o $@ $< -module .

$(OBJ): $(SRC)
    $(PAR_FC) $(FLAGS) -c -o $@ $<


.PHONY: clean
clean:
    $(RM) -f $(EXE) *.o *.mod
\end{lstlisting}
\begin{lstlisting}[language={bash},frame=single,numbers=left,numberstyle=\tiny,basicstyle=\footnotesize,stepnumber=1,breaklines=true,caption={Batch file},label=lst:batch]
#!/bin/bash -l
#SBATCH -J L80_2N3N4N
#SBATCH -N 100
#SBATCH --ntasks-per-node=24
#SBATCH --time=1:00:00
#SBATCH -A hk4d3nn
#SBATCH --output="pi_4d_2n3n4n.out"
#SBATCH --error="pi_4d_2n3n4n.err"

cd $SLURM_SUBMIT_DIR

module add plgrid/tools/impi
export FORT_BUFFERED=yes

mpiexec -env I_MPI_FABRICS shm:dapl -env I_MPI_FABRICS_LIST dapl -env I_MPI_DAPL_UD=enable -env MPIEXEC_PREFIX_DEFAULT=out ./pi_4d_2n3n4n.exe 80 0.030 0.034 0.0001 10000
\end{lstlisting}
\end{widetext}
\end{document}